\documentclass[12pt]{article}
\usepackage[mathscr]{eucal}
\usepackage{epsfig,amsfonts}
\usepackage[fleqn]{amsmath}
\usepackage{amssymb}
\usepackage{graphicx}
\usepackage{hhline}
\usepackage{cite}
\usepackage{mathrsfs}
\usepackage{hyperref}
\usepackage{verbatim}
\usepackage{stmaryrd}
\usepackage{lipsum}
\usepackage{amsfonts}
\usepackage{amscd}
\usepackage{bbm}

\makeatletter \@addtoreset{equation}{section} \makeatother

\def\gl11{\mathfrak{gl}(1|1)}
\def\afgl11{\mathfrak{gl}(1|1)^{\vee }}
\def\Uhgl11{U_h(\gl11)}
\newtheorem{theorem}{Theorem}
\newtheorem{lemma}{Lemma}
\newtheorem{proposition}{Proposition}
\newtheorem{corol}{Corollary}

\begin{document}

{\Large \textbf{The Drinfeld-Kohno theorem for the superalgebra $\mathfrak{gl%
}(1|1)$} }

\begin{center}
\vspace{1cm}

{\large Andrei Babichenko\footnote{{\large babichenkoandrei@gmail.com}} }

\vspace{1.cm}

Department of Mathematics, The Weizmann Institute of Science,\\[0pt]
Rehovot 76100, Israel \\[0pt]
\vspace{.2cm}
\end{center}

\begin{center}
\centerline{\bf Abstract} \vspace{.8cm}
\parbox{15.5cm}
{ We revisit the derivation of Knizhnik-Zamolodchikov equations in the case of  non-semisimple categories of modules of a superalgebra in the case of the generic affine level and representations parameters. A proof of existence of asymptotic solutions and their properties for the superalgebra $\gl11$ gives a basis for the proof of existence associator which satisfy braided tensor categories requirements. Braided tensor category structure of $\Uhgl11$ quantum algebra calculated, and the tensor product ring is shown to be isomorphic to $\gl11$ ring, for the same generic relations between the level and parameters of modules. We review the proof of Drinfeld-Kohno theorem for non-semisimple category of modules suggested by Geer \cite{Gee} and show that it remains valid for the superalgebra $\gl11$. Examples of logarithmic solutions of KZ equations are also presented. }
\end{center}


\section{Introduction}

Drinfeld - Kohno (DK) theorem \cite{Dr1} - \cite{Ko} states braided tensor
equivalence between seemingly different categories of modules: on the one
hand, quasitriangular quasi-Hopf universal enveloping algebra modules
associated to a simple Lie algebra $\mathfrak{g}$ with associator and
braiding defined through Knizhnik-Zamolodchikov (KZ) equation with quantum
deformation parameter $h$, and on the other hand -- of modules of
quasitriangular Hopf $h$-quantized universal enveloping algebra associated
to $\mathfrak{g}$. By this equivalence the quantization parameter $h$ of the
latter algebra corresponds in the former to deformation parameter of
associator which arises as a monodromy of KZ equation solutions associated
with the chosen representations category. This theorem was proved by
Drinfeld using series expansion in $h$ around zero, and is valid for generic
values of this parameter, with excluded specific rational values. Later on,
in the seminal series of papers \cite{KL1} - \cite{KL4}, this equivalence
were addressed more generally by Kazhdan and Lusztig. In \ \cite{KL1}, \cite%
{KL2} they showed that when $l\notin
\mathbb{Q}
$ or when $l\in
\mathbb{Q}
$ but $l<-h^{\vee }$, where $h^{\vee }$ is the dual Coxeter number of $%
\mathfrak{g}$, a certain category of affine level $l$ Lie algebra $\widehat{%
\mathfrak{g}}$ modules has a natural braided tensor category structure, and
they proved rigidity for most of these tensor categories in \cite{KL4}.
Kazhdan - Lusztig construction was then extended by Finkelberg to rational
positive level categories of affine modules in \cite{Fin}. In this affine
algebraic context the KZ equation appears naturally, and Kazhdan and Lusztig
have proved \cite{KL3}, \cite{KL4} braided tensor equivalence of their
category at level $l$ to the category of finite-dimensional modules over the
quantum group $U_{q}(\mathfrak{g})$ where $q=e^{\frac{i\pi }{m((l+h^{\vee })}%
}$, $m$ is the ratio of the squared length of the long roots of $\mathfrak{g}
$ to the squared length of the short roots.

The interest to this equivalence of representation categories was renewed in
the context of attempts to understand representation theory of logarithmic
conformal field theories \cite{FSem1}, \cite{FSem2} or of logarithmic vertex
operator algebras (VOA) - their mathematically rigorous incarnation (see
e.g. \cite{HuRev} and references therein in for mathematically oriented, and
\cite{CRbeyond} for physically oriented reviews). One of the main
ingredients which differ logarithmic VOA from rational ones is essential
role played by reducible but indecomposable modules. The current
understanding of representation theory of logarithmic VOA is far from being
complete. Since set of intertwiner operators of VOA satisfy KZ equations,
analogs of DK theorem, and especially its extension to all the values of
deformation parameter, can add to understanding of the representation theory
of logarithmic VOAs. The VOA related to the affine Lie superalgebra $\hat{\mathfrak{gl}}(1|1)$
is one of the archetypical examples of logarithmic VOAs \cite{CRarhy}. This
motivates to start from DK theorem for this algebra for suitable category of
representations, for generic values of the affine level, with a hope to
extend analysis of this example beyond the scope of generic values, with
further extension to logarithmic VOAs. The description of the category of
modules we consider and restrictions on their parameters corresponding to
situation of generic level (deformation parameter) will be given below.

Of course, the question about DK theorem for superalgebras was addressed
before. It turns out that direct copy of Drinfeld's proof of DK theorem for
Lie superalgebras is impossible because of the obstacles explained in
particular in \cite{Gee}. Nevertheless the author succeeded to prove DK
theorem for the classical superalgebras applying Etingof-Kazhdan approach to
quantization \cite{EK1} - \cite{EK3} as a bridge for tensor equivalence. We
refer to \cite{Gee} and references therein for details, which will be
reviewed below.

The main object which makes the equivalence of categories explicit is the
twist $\mathcal{F}$. Its explicit, non perturbative in $h$ construction in
the case of simple Lie algebras is difficult. Some attempt of such explicit
construction for simple Lie algebras known to us, without proofs that the
constructions indeed implement full braided tensor equivalence, is \cite%
{Engel}. It is based on basis dependent fundamental representations
projectors of simple Lie algebras. Our way of rigorous proof of tensor
equivalence is a repeat of the proof of Geer with a trivial argumentation
why it works for the case of non-semisimple Lie superalgebra $\mathfrak{gl}(1|1)$, which
formally not in the list of superalgebras he considered.

The main result of the paper is the Theorem \ref{thm7}. It claims that for
the superalgebra $\mathfrak{gl}(1|1)$ two non-semisimple categories of
modules are braided tensor equivalent. The first category is the Drinfeld
category $\mathcal{D}$ with (equivalence classes of) the typical $\mathcal{T}%
_{e,n}$, atypical $\mathcal{A}_{n}$, and projective $\mathcal{P}_{n}$
modules as objects, such that the parameters $e_{i}$ satisfy $e_{i}/\kappa
\notin
\mathbb{Z}
$ and $(e_{i}+e_{j})/\kappa \notin
\mathbb{Z}
\backslash \{0\}$ for any pair of typical modules. The second category is
the tensor category $\mathcal{C}_{\kappa }$ of corresponding modules $%
\mathcal{T}_{e,n}^{\kappa }$, $\mathcal{A}_{n}^{\kappa }$, $\mathcal{P}%
_{n}^{\kappa }$ of quantum group $U_{\kappa }(\mathfrak{gl}(1|1))$

The paper is organized as follows. In the next Section \ref{section2} we
review the main steps of derivation of KZ equations, first in operator form
for intertwining operators, then -- for correlation functions of
intertwiners. There is almost no difference in it compared to Lie algebra
case when non-semisimple finite dimensional modules are included. In the
Section \ref{section3} we define Drinfeld category $\mathcal{D}$ for any Lie
(super)algebra, and its tensor ring structure in the $\mathfrak{gl}(1|1)$ case for three
types of $\mathfrak{gl}(1|1)$-modules. The main part of this section is the proof of
existence of associator in the $\mathfrak{gl}(1|1)$ case with its standard properties, as
well as the braiding. The Section \ref{section4} defines the category $%
\mathcal{C}_{\kappa }$ of corresponding $U_h(\mathfrak{gl}(1|1))$ quantum group modules with
its tensor product ring and other braided tensor category structures. The
Section \ref{section5} reviews different aspects of proof of equivalence of
the two categories of modules. Some perspectives of continuation of this
research is summarized in the Section \ref{section6}. Many technical
details, such as bases of the representations, solutions of KZ equations,
their asymptotic needed for the proof of associator existence are collected
in the Appendix A \ref{appA}. Similar technical information about the
quantum group side, including the proof of the tensor product ring structure
of the modules in specified bases one can find in the Appendix B \ref{appB}.

For the rest of the paper we make an important remark:

\textit{The proofs of statements and theorems cited below as known do not
use the fact of algebra semisimplicity or semisimplicity of the category of
its modules under consideration. The cases where it requires different
proofs or leads to different results (like as in analysis of asymptotic
solutions of KZ equations) are considered in details. Modifications of
proofs related to the fact that we deal with superalgebra are trivial and do
not change the cited statements of known theorems. The only needed
modifications is in definition of }$%
\mathbb{Z}
_{2}$ \textit{graded commutator}%
\begin{equation*}
\lbrack A,B]=AB-(-1)^{p(A)p(B)}BA
\end{equation*}%
\textit{and the manipulations with tensor products}%
\begin{equation*}
(A\otimes B)(a\otimes b)=(-1)^{p(B)p(a)}Aa\otimes Bb
\end{equation*}%
\textit{where }$p(x)$ \textit{is the parity of the object }$x$. \textit{An
exception from this general rule appears in tensor product decomposition of }%
$%
\mathbb{Z}
_{2}$\textit{\ graded modules which sometimes involve parity reverse
operator. (It will be explained in the proper cases below)}

\textbf{Acknowledgements. }The author is thankful to I.Scherbak for
clarifying explanations related to asymptotic solutions of KZ equations, and
especially grateful to P.Etingof for many valuable stimulating discussions.


\section{Generic $\protect\kappa $ KZ equation}

\label{section2}

Below we recall standard derivation of operator KZ equation for intertwiners
of any affine algebra $\widehat{\mathfrak{g}}$ with some remarks specifying
the super case, for affinization of any category of finite dimensional $%
\mathfrak{g}$-modules (possibly indecomposable) at generic $\kappa $. By $%
\kappa $ we denote the inverse quantization parameter discussed above $%
\kappa =h^{-1}=h^{\vee }+k$, $h^{\vee }$ is dual Coxeter number and $\ k$ is
the level of affine (super)algebra $\mathfrak{g}$. We also recall standard
derivation of KZ equations for correlation functions. The fact that some of
modules are indecomposable doesn't hamper to repeat the standard steps of
derivation \textit{for generic }$\kappa $ (see \cite{EFK} Lecture 3 for a
review). In the case of $\mathfrak{gl}(1|1)$ \ we have $h^{\vee }=0$ and
generic means generic values of $k$ which will be specified below.

\subsection{Intertwining operators}

\label{subIntertwin}

We start from\ formulation of affine intertwiners of Tsuchia and Kanie \cite%
{TK}, summarized in \cite{EFK}, Lecture 3, generalizing it to the
non-semisimple $\mathfrak{g}$-modules. Let $\mathfrak{g}$ be a simple Lie
(super)algebra over $%
\mathbb{C}
$. Let $M_{p}$ be a \textit{finite dimensional} indecomposable (possibly
reducible) $\mathfrak{g}$-module, $p$ - some set of parameters which
characterise the module. The module is weight: for any homogeneous vector $%
u\in M_{p}$, $\mathfrak{h}u=\lambda _{u}u$ for some $\lambda _{u}\in
\mathbb{C}
$. We assume that Casimir element $\Omega $ of $U(\mathfrak{g)}$ can act non
diagonally and we decompose $\Omega =C_{d}+C_{nil}$ where $C_{d}$ acts
diagonally with the same eigenvalue $\lambda _{p}$ on all the vectors of the
module, and $C_{nil}$ is a nilpotent part of non-diagonal action: $%
(C_{nil})^{n}=0$ for some non negative integer $n$. In the case of non-super Lie algebras $M_{p}$
is assumed to be a highest weight module and $C_{nil}=0$. We relax this
requirement.

In what follows all commutations and tensor products are understood as $%
\mathbb{Z}
_{2}$ graded for the case of superalgebras. We recall some notations and
definitions related to affine Lie (super)algebraic modules. We consider
triangular decomposition with respect to $%
\mathbb{Z}
$-grading $\widehat{\mathfrak{g}}=\widehat{\mathfrak{g}}_{<0}\mathfrak{%
\oplus }\widehat{\mathfrak{g}}_{0}\mathfrak{\oplus }\widehat{\mathfrak{g}}%
_{>0}$, and induced $\widehat{\mathfrak{g}}$-modules $M_{p,k}=Ind_{\widehat{%
\mathfrak{g}}_{\geq 0}}^{\widehat{\mathfrak{g}}}M_{p}$, \textit{for generic }%
$k$ , where the action of $\widehat{\mathfrak{g}}_{>0}=\mathfrak{g\otimes t%
\mathbb{C}
\lbrack t]}$ is trivial, and the action of $\widehat{\mathfrak{g}}_{0}=%
\mathfrak{g\oplus }k\mathfrak{%
\mathbb{C}
}$ is such that it is isomorphic to the action of $\mathfrak{g}$ for the
first summand, and is multiplication by $k$ - for the second. The modules $%
M_{p,k}$ admit $%
\mathbb{Z}
$-grading
\begin{equation*}
M_{p,k}=\bigoplus\limits_{n\geq 0}M_{p,k}[-n]
\end{equation*}%
and $M_{p,k}[-n]$ is the eigenspace of affine Lie (super)algebra derivation $%
d$ with the eigenvalue $-n-\Delta _{p,k}$, $\Delta _{p,k}$ is the conformal
dimension. The $M_{p,k}[0]$ is naturally a $\mathfrak{g}$-module isomorphic
to $M_{p}$.

Recall that for simple Lie algebras we can restrict and define as generic $%
\kappa =k+h^{\vee }$ such that $k\notin
\mathbb{Q}
$. In our superalgebra case we mean by generic $k\in
\mathbb{C}
$ restricted by suitable constraints dependent on modules parameters, such
that they guarantee that if $M_{p}$ is irreducible then $M_{p,k}$ is also
irreducible, and if $M_{p}$ is of finite length with composition factors $%
L_{p_{i}}$ then $M_{p,k}$ is also of finite length with the composition
factors $L_{p_{i},\kappa }=Ind_{\widehat{\mathfrak{g}}_{\geq 0}}^{\widehat{%
\mathfrak{g}}}L_{p_{i}}$. We will see below what restrictions on the level $%
k $ for the affine $\mathfrak{gl}(1|1)_k$ case it implies.

Another important kind of $\widehat{\mathfrak{g}}$-modules we need in order
to define affine intertwiner is \textit{evaluation module. }Let $M_{p}$ be a
$\mathfrak{g}$-module which admits weight decomposition with finite
dimensional weight spaces. For a non zero complex number $z$ and for any
element $x\otimes P(t)$ of $\mathfrak{g}\otimes \mathfrak{%
\mathbb{C}
}[t,t^{-1}]$ where $P$ is a polynom, we define its action $x\otimes
P(t)\cdot u=P(z)xu$, where $u\in M_{p}$, and the central element of $%
\widehat{\mathfrak{g}}$ acts trivially. After that one can extend such
action on the whole $\widehat{\mathfrak{g}}$ by replacing the space of
action by a bigger one $M_{p}\otimes z^{-\Delta }\mathfrak{%
\mathbb{C}
}[z,z^{-1}]\footnote{%
Strictly speaking it requires to consider $z$ as a formal variable $z$ and
restoration of status of complex variable requires subtle procedure worked
out in vertex operator algebras formalism by Huang and Lepowsky, and later
by Huang, Lepowsky and Zhang for logarithmic vertex operator algebras (see
the list of references in \cite{HuRev})}$, With a standard extension of $%
\widehat{\mathfrak{g}}$ to $\widetilde{\mathfrak{g}}$ by derivation $d$, we
define the action of the affine derivation as $d=z\frac{d}{dz}$. \ Here $%
\Delta \in
\mathbb{C}
$ is specified as module dependent parameter (its conformal dimension). Now
we have the generating function for $%
\mathbb{Z}
$-graded modes $v[n]=v\otimes t^{n}\in M_{p,k}$ : $v(z)=\sum_{n\in
\mathbb{Z}
}v[n]z^{-\Delta -n}$. We denote the vector space of such objects as $%
M_{p}(z) $.

We are equipped now with all necessary ingredients for construction of
affine intertwiner. Assume we can classify all $\mathfrak{g}$-homomorphisms
of the form $\varphi :M_{p_{1}}\rightarrow M_{p_{0}}\otimes M_{p}$, $g\in
\mathfrak{g}$. We call them $\mathfrak{g}$-intertwiners and we want to lift
them to $\widehat{\mathfrak{g}}$-intertwiners $\Phi $. Our consideration
will be restricted to the special class of intertwiners which preserve the
superalgebra $%
\mathbb{Z}
_{2}$ grading. We define a $\widehat{\mathfrak{g}}$-intertwiner as a
homomorphism $\Phi :M_{p_{1},k}\rightarrow M_{p_{0},k}\widehat{\otimes }%
M_{p}(z)$ which satisfies
\begin{equation*}
\Phi (z)x[n]=(x[n]\otimes 1+z^{n}\cdot 1\otimes x)\Phi (z)
\end{equation*}%
where $z$ is a non zero complex number. Here $\widehat{\otimes }$ is
understood as a completed tensor product consisting of infinite sums of
tensor products of homogeneous vectors.

Recall that in the case of simple non-super algebras, for the picture of
complete braided tensor category (BTC) structure, it is enough to consider
affinization of finite dimensional highest weight $\mathfrak{g}$-modules
(sometimes called Weyl modules), and evaluation modules. We will do the same
for superalgebras relaxing the condition that we affinize and build
evaluation modules over highest weight irreducible modules: the $\mathfrak{g}
$-modules are not necessarily highest weight, and may be reducible but
indecomposable. Almost all the steps of intertwiners construction can be
copied from the non-super case. In particular the following assertion can be
proved as a slight generalization to the superalgebra case of the Theorem
3.1.1 of \cite{EFK}, which was originally proved in \cite{TK} in $\widehat{%
\mathfrak{sl}}(2)$ case (Theorem 1, followed from Proposition 2.1 and
Theorem 2.2)

\begin{proposition}
\label{pr0}Let $\varphi :M_{p_{1}}\rightarrow
M_{p_{0}}\otimes M_{p}$ be a $\mathfrak{g}$-homomorphism. \ Then for generic
$k$ there exists a unique $\widehat{\mathfrak{g}}$-intertwiner $\Phi
(z):M_{p_{1},k}\rightarrow M_{p_{0},k}\widehat{\otimes }M_{p}(z)$ such that
for every vector $v\in M_{p_{1},k}[0]\simeq $ $M_{p_{1}}$the zero degree
component of $\Phi (z)v$ is equal to $\varphi v$.
\end{proposition}

We recall here the sketch of the proof\textit{.} Because of the annihilation
condition for $w\in M_{p_{1},k}[0]\simeq M_{p_{1}}$ by $\mathfrak{g\otimes t%
\mathbb{C}
\lbrack t]}$ we have
\begin{equation}
\Phi (z)w\in \left( M_{p_{0},k}\widehat{\otimes }M_{p}(z)\right) ^{\mathfrak{%
g}\otimes t%
\mathbb{C}
\lbrack t]}=Hom_{\mathfrak{g}\otimes t%
\mathbb{C}
\lbrack t]}\left( M_{p_{0},k}^{\ast },M_{p}(z)\right)  \label{n1}
\end{equation}%
The contragredient module $M_{p_{0},k}^{\ast }$ is freely generated over $%
\mathfrak{g\otimes t%
\mathbb{C}
\lbrack t]}$ \textit{for generic }$k$, therefore the restriction map
\begin{equation*}
Hom_{\mathfrak{g}\otimes t%
\mathbb{C}
\lbrack t]}\left( M_{p_{0},k}^{\ast },M_{p}(z)\right) \rightarrow Hom_{%
\mathbb{C}
}\left( M_{p_{0}}^{\ast },M_{p}\right) \simeq (\Pi )M_{p_{0}}\otimes M_{p}
\end{equation*}%
is an isomorphism. ($\Pi $ means the parity inversion which sometimes is
necessary in superalgebra case, see below.) Therefore $\Phi (z)w$ is
uniquely defined by its zero grade component.

Then we define the homomorphic action of $\Phi (z)$ on any $Xw=u\in
M_{p_{1},k}$, where $X\in U(\widehat{\mathfrak{g}})$, $X=\prod%
\limits_{i}x_{n_{i}}^{(i)}$, is written in PBW basis, where $x^{(i)}\in
\mathfrak{g}$ and $n_{i}\in
\mathbb{Z}
$. We define it inductively over each factor $x_{n_{i}}^{(i)}$ of $X$ by
\begin{equation}
\Phi (z)x_{n_{i}}^{(i)}=[x_{n_{i}}^{(i)}\otimes 1+z^{n}(1\otimes
x_{n_{i}}^{(i)})]\Phi (z),\;n<0  \label{n2}
\end{equation}%
It defines an $\widehat{\mathfrak{g}}$-intertwiner uniquely with an obvious
property that it is a lifting of the $\mathfrak{g}$-intertwiner $\varphi $.

A subtle point of the above definition is that some more general and
rigorous construction is needed in order to treat $z$ as a genuine complex
variable and not just formal variable. This construction was elaborated in
the seminal series of papers by Huang and Lepowsky - see the footnote above
- in the framework of vertex operator algebras (VOA). In particular more
general intertwiners of the form
\begin{equation*}
\mathcal{Y}(\_,z):W_{1}\rightarrow Hom(W_{2},W_{3})\{z\}[\log z]
\end{equation*}%
are usually needed in logarithmic vertex operator superalgebra (VOSA) $V$
case, where $W_{i}$ are some $V$-modules. It is precisely relevant for our $%
\mathfrak{gl}(1|1)$ case, but we will continue to use the definition of
intertwiners described above, without use of VOSA language. Despite the lack
of proper rigorously we will continue to treat $z$ in our approach as
complex variable, defining, where it is needed a branch of multivalued
functions. In particular for what follows we chose $\ln z=\ln
|z|+iArg(z),-\pi <Arg(z)\leq \pi $.

An important remark is in order here. Recently, when a preliminary version
of this paper was finished, an important progress was achieved in
understanding of braided tensor category structure of the $\mathfrak{gl}%
(1|1) $ VOSA \cite{CMY}.

The next standard step is to extend this $\widehat{\mathfrak{g}}$%
-homomorphism to $\widetilde{\mathfrak{g}}$-homomorphism, where $\widetilde{%
\mathfrak{g}}$ is the standard extension of $\widehat{\mathfrak{g}}$ by
affine derivation $d=-L_{0}$, with $L_{m}$ defined by Sugawara construction.
\begin{equation}
L_{m}=\frac{1}{2\left( k+h^{\vee}\right) }\sum_{a,b}\sum_{n\in%
\mathbb{Z}
}B_{ab}^{-1}:J_{n}^{a}J_{m-n}^{b}:  \label{n4}
\end{equation}
where $h^{\vee}$ is a dual Coxeter number of (super)algebra\footnote{%
This construction can be modified in the case of non semisimple
(super)algebra. It acts as a scalar on simple modules, but sometimes acts
non diagonally on indecomposables, as for example in the case of $\widehat{gl%
}(1|1)$.}, and $B$ - $\mathfrak{g}$-invariant (super)symmetric
non-degenerated bilinear form. If we want to extend the intertwining
homomorphism $\Phi(z)$ defined in Proposition \ref{pr0} to $\widetilde{\mathfrak{g}}$-homomorphism we
have to twist it. We define two twisted intertwiners: for $w\in M_{p_{1},k}$

\begin{align}
\widehat{\Phi }^{g}(z)w& =(z^{L_{0}}\otimes z^{L_{0}})\left( z^{-L_{0}}\Phi
(z)wz^{L_{0}}\right) (z^{-L_{0}}\otimes z^{-L_{0}}),  \label{n3} \\
\widetilde{\Phi }^{g}(z)w& =(z^{L_{0}}\otimes 1)\left( z^{-L_{0}}\Phi
(z)wz^{L_{0}}\right) (z^{-L_{0}}\otimes 1)  \label{n31}
\end{align}%
They remain intertwiners with image in
\begin{equation}
z^{-L_{0}}M_{p_{0},k}z^{L_{0}}\widehat{\otimes }%
z^{-L_{0}}M_{p}z^{L_{0}}[z,z^{-1}]  \notag
\end{equation}%
and
\begin{equation}
z^{-L_{0}}M_{p_{0},k}z^{L_{0}}\widehat{\otimes }M_{p}[z,z^{-1}]  \notag
\end{equation}%
respectively. In the case of irreducible highest weight modules $M_{p_{i}}$
with highest weight $p_{i}$ these twists reduce to the standard scalar
factors twists $\widehat{\Phi }^{g}(z)=\sum_{n}\Phi(n)z^{-n-\Delta }$, $%
\Delta =\Delta (p_{1})-\Delta (p_{0})-\Delta (p)$, and the same for $%
\widetilde{\Phi }^{g}$ with $\Delta (p_{1})-\Delta (p_{0})$, where $\Delta
_{i}=\frac{\langle p_{i},p_{i}+2\rho \rangle }{2(k+h^{\vee })}$. (The factor
$z^{\Delta (p_{0})+\Delta (p)}$ is moved to the definition of $\widehat{\Phi
}^{g}(z)$ by the first and the last parenthesis factors.)

For the restricted dual $M_{p}^{\ast }$ and its evaluation module $%
M_{p}^{\ast }(z)\cong (M_{p}(z))^{\ast }$ which are assumed to be well
defined, we can take any vector $u\in M_{p}^{\ast }$, define $\widehat{\Phi }%
_{u}^{g}(z)w=\langle 1\otimes u,\widehat{\Phi }^{g}(z)w\rangle ,w\in
M_{p_{1},k}$ and regard it as an operator $\widehat{\Phi }%
_{u}^{g}(z):M_{p_{1},k}\rightarrow M_{p_{0},k}$. Then the proof of the
theorem \cite{KZ}, \cite{FR} Theorem 2.1, about the operator form of KZ
equation which says that
\begin{equation}
(k+h^{\vee })\frac{d}{dz}\widehat{\Phi }_{u}^{g}(z)=\sum_{a\in B}:J_{a}(z)%
\widehat{\Phi }_{au}^{g}(z):  \label{n8}
\end{equation}%
(summation is over the basis $B$ of $\mathfrak{g}$) generalizes to the case
of indecomposable modules $M_{p_{i}}$ actually without changes. Recall the
proof.

Obviously the intertwining relation (\ref{n2}) is satisfied for $\widehat{%
\Phi }^{g}(z)$ as well. Applying contravariant bilinear form in the space $%
M_{p}$ this relation can be written as
\begin{equation*}
\lbrack \widehat{\Phi }_{u}^{g}(z),x[n]]=z^{n}\widehat{\Phi }_{xu}^{g}(z)
\end{equation*}%
If we introduce currents $J_{x}^{\pm }(z)$ for any algebra element $x$
\begin{equation*}
J_{x}(z)=J_{x}^{+}(z)-J_{x}^{-}(z),\ J_{x}^{+}(z)=\sum_{n<0}x[n]z^{-n-1},\
J_{x}^{-}(z)=-\sum_{n\geq 0}x[n]z^{-n-1}
\end{equation*}%
then in terms of these currents the last intertwining property takes the form%
\begin{equation}
\lbrack J_{x}^{\pm }(\zeta ),\widetilde{\Phi }_{u}^{g}(z)]=\frac{1}{z-\zeta }%
\widetilde{\Phi }_{xu}^{g}(z)  \label{crcur}
\end{equation}%
(plus sign corresponds to $|\zeta |<|z|$, and minus sign -- to $|\zeta |>|z|$%
). Now we write the $d$-invariance property of $\widetilde{\Phi }_{u}^{g}(z)$%
:%
\begin{equation*}
z\frac{d}{dz}\widetilde{\Phi }_{u}^{g}(z)=-[d,\widetilde{\Phi }_{u}^{g}(z)]
\end{equation*}%
which is the same as
\begin{equation*}
z\frac{d}{dz}\widehat{\Phi }_{u}^{g}(z)=-[d,\widehat{\Phi }_{u}^{g}(z)]+z%
\frac{d}{dz}(1\otimes z^{-L_{0}})\widehat{\Phi }_{u}^{g}(z)(1\otimes
z^{L_{0}})
\end{equation*}%
We can continue by Sugawara construction
\begin{align*}
& \frac{B_{a,b}^{-1}}{2(k+h^{\vee })}\left( \sum_{n\leq 0}[J^{a}[n]J^{b}[-n],%
\widehat{\Phi }_{u}^{g}(z)]+\sum_{n>0}[J^{a}[-n]J^{b}[n],\widehat{\Phi }%
_{u}^{g}(z)]\right) + \\
&z\frac{d}{dz}(1\otimes z^{-L_{0}})\widehat{\Phi }%
_{u}^{g}(z)(1\otimes z^{L_{0}})= \\
& \frac{B_{a,b}^{-1}}{2(k+h^{\vee })}\{2zJ_{b}^{+}(z)\widehat{\Phi }%
_{au}^{g}(z)-2z\widehat{\Phi }_{bu}^{g}(z)J_{a}^{-}(z)+J_{b}^{+}[0]\widehat{%
\Phi }_{au}^{g}(z)-\widehat{\Phi }_{bu}^{g}(z)J_{a}^{-}[0]\} \\
& +z\frac{d}{dz}(1\otimes z^{-L_{0}})\widehat{\Phi }_{u}^{g}(z)(1\otimes
z^{L_{0}})= \\
& \frac{B_{a,b}^{-1}}{k+h^{\vee }}:J_{a}(z)\widehat{\Phi }_{bu}^{g}(z):+%
\frac{B_{a,b}^{-1}}{2(k+h^{\vee })}(J_{b}^{+}[0]\widehat{\Phi }_{au}^{g}(z)-%
\widehat{\Phi }_{bu}^{g}(z)J_{a}^{-}[0])+ \\
&z\frac{d}{dz}(1\otimes z^{-L_{0}})%
\widehat{\Phi }_{u}^{g}(z)(1\otimes z^{L_{0}})
\end{align*}%
The last two terms cancel because they can be written as
\begin{equation*}
\frac{1}{2(k+h^{\vee })}\widehat{\Phi }_{Cu}^{g}(z)-\Delta (p)\widehat{\Phi }%
_{u}^{g}(z)
\end{equation*}%
where $C=B_{a,b}^{-1}J_{a}J_{b}$ is a Casimir element of the algebra $%
\mathfrak{g}$, (not to be confused with tensor Casimir defined in (\ref{n15}%
). This completes the proof.

Concluding this section about systematic definition of intertwining
operators for affine Lie (super)algebra we can illustrate an important
difference of a non-semisimple case from a semisimple one. Suppose we have a
finite dimensional $\mathfrak{g}$-module with non-semisimple action of the
Casimir element $C$ which we can represent as $C=\lambda I+C_{nil}$, where $%
I $ acts as identity and $C_{nil}$ acts nilpotently: $C_{nil}^{n}=0$ on each
vector of the module. Then the action of $z^{aC}$ where $z,\lambda \in
\mathbb{C}
$, on any vector $w$ of the module can be written as%
\begin{equation}
z^{C}w=z^{\lambda }\sum_{m=1}^{n-1}\frac{(\ln z)^{m}}{m!}C_{nil}^{m}w
\label{n6}
\end{equation}%
leading to a presence of logarithms (with a choice of a branch), which
necessarily arises with non-semisimplicity in logarithmic vertex operator
algebras and in logarithmic conformal field theories. A modification of this
example for the operator like $z^{\Omega _{ij}}$ will appear below where the
Casimir $C$ is replaced by quantum Casimir $\Omega _{ij}$, see (\ref{n15})
below.

\subsection{KZ equation for correlation functions}

\label{subCorFun}

The way of derivation of correlation functions KZ equation from operator KZ
equation (\ref{n8}) first appeared in the seminal paper of Knizhnik and
Zamolodchikov \cite{KZ}. Later it was derived more rigorously in \cite{FR}
and in textbooks like e.g. \cite{EFK}, Sect. 3.4. Possible
non-semisimplicity of Lie (super)algebra $\mathfrak{g}$-modules doesn't lead
to serious modifications in the derivation. We again recall the main steps
of it.

In order to define correlation function consider the modules $%
M_{q_{i},k},i=1,...,N$, and $M_{p_{i}}$, $i=1,...,N+1$. Let $\widehat{\Phi }%
^{g_{i}}(z_{i}):M_{p_{i},k}\rightarrow M_{p_{i-1},k}\widehat{\otimes }%
M_{q_{i}}[z_{i}^{\pm 1}]$ be an intertwiner as explained above, where $%
M_{q_{i}}(z_{i})$ is evaluation module. We consider the homomorphism%
\begin{eqnarray}
\Psi (z_{1},...,z_{N})=\left( \widehat{\Phi }^{g_{1}}(z_{1})\otimes
1...\otimes 1\right) ...\left( 1...\otimes \widehat{\Phi }%
^{g_{N-1}}(z_{N-1})\otimes 1\right) \times \nonumber\\
\left( 1\otimes ...\otimes \widehat{\Phi}^{g_{N}}(z_{N})\right)  \label{n10}
\end{eqnarray}%
that maps $M_{p_{N},k}\rightarrow M_{p_{0},k}\widehat{\otimes }M_{q_{1}}%
\widehat{\otimes }...\widehat{\otimes }M_{q_{N}}$.

This formula for homomorphism makes sense at least being understood as
formal power series in $z_{1},z_{2},...,z_{N}$ and their logarithms.

Consider a subspace of weight $\lambda_{N}$ of $M_{p_{N},k}[0]$, and
subspace of weight $-\lambda_{0}$ of $M_{p_{0},k}^{\ast}[0]$. The object $%
\Psi (z_{1},...,z_{N})|\lambda_{N}\rangle$ takes values in the space $%
M_{q_{1}}\otimes M_{q_{2}}...\otimes M_{q_{N}}\otimes M_{p_{0}}$. We can
take a projection of it onto finite dimensional invariant subspace of the
weight $\lambda_{N}-\lambda_{0}$ in the $M_{p_{0}}$ component of it $%
V=\left( M_{q_{1}}\otimes M_{q_{2}}...\otimes M_{q_{N}}\right)
^{\lambda_{N}-\lambda_{0}}$. If we take $\lambda_{N}=\lambda_{0}$ then we
get the $\mathfrak{g}$ invariant subspace $V^{\mathfrak{g}}$. This sort of
projection of $\Psi$ on such a subspace, with some chosen $u_{N+1}\in
M_{p_{N},k}[0]$, $u_{0}\in$ $M_{p_{0},k}[0]$, is called a \textit{%
correlation function}
\begin{align}
\psi(z_{1},...z_{N}) & =\langle u_{0},\Psi(z_{1},...z_{N})u_{N+1}\rangle
\label{n111} \\
\psi(z_{1},...z_{N}) & \in\left( M_{q_{1}}\otimes M_{q_{2}}...\otimes
M_{q_{N}}\right) ^{\lambda_{N}-\lambda_{0}}  \notag
\end{align}
(the vector $\langle u_{0}|\in M_{p_{0},k}^{\ast}[0]$ ) Taking into account
the remark (\ref{n6}) we can say that $\psi$ here is defined as a formal
power series: it belongs to \\ $\prod
\limits_{i}z_{i}^{-\Delta(p_{i})+\Delta(p_{i-1})+\Delta(q_{i})}(\ln\frac{%
z_{i}}{z_{i-1}})^{n_{i}}%
\mathbb{C}
\lbrack\lbrack\frac{z_{2}}{z_{1}},...\frac{z_{N}}{z_{N-1}}]]$. Equivalently
one can define correlation function as $%
\mathbb{C}
$-valued if choosing $u_{i}\in M_{q_{i}},i=1,...,N$, we define
\begin{equation}
\psi_{u_{1},...,u_{N+1}}(z_{1},...z_{N})=\langle u_{0},\widehat{\Phi}%
_{u_{1}}^{g_{1}}(z_{1})...\widehat{\Phi}_{u_{N}}^{g_{N}}(z_{N})u_{N+1}%
\rangle\in%
\mathbb{C}
\label{n13}
\end{equation}
In particular one can take $M_{p_{0},k}=M_{p_{N},k}$ to be the scalar
representation $M_{0}$, i.e. $M_{p_{0},k},M_{p_{N},k}$ -- induced vacuum
modules with the zero grade vector $u_{0}$, and define $V$-valued
correlation function.
\begin{equation}
\phi(z_{1},...z_{N})=\langle u_{0},\Psi(z_{1},...,z_{N})u_{0}\rangle
\label{n12}
\end{equation}
Then $\phi(z_{1},...z_{N})\in V^{\mathfrak{g}}$.\footnote{%
In the super algebras case it sometimes happens that a scalar representation
appears only as a (part of) atypical module. By general tensor category
"ideology" atypical modules should be replaced by their projective covers.
But even then there is a "bottom" vector $u_{N+1}$ in it satisfying $%
\mathfrak{g}u_{N+1}=0$.}

The main theorem proved in \cite{KZ} for simple highest weight modules of
(non super) algebra, claims the KZ equation on (\ref{n111}) in the form
\begin{equation}
(k+h^{\vee })\partial _{i}\psi =\left( \sum_{j\neq i=1}^{N}\frac{\Omega _{ij}%
}{z_{i}-z_{j}}+\frac{\Omega _{i,N+1}}{z_{N}}\right) \psi ,\;i=1,...,N+1
\label{n141}
\end{equation}%
Equivalent form of KZ equation can be obtained by adding one more formal
variable $z_{N+1}$ to the function \ $\psi (z_{1},...z_{N})=\psi
(z_{1}-z_{N+1},...z_{N}-z_{N+1})$, giving
\begin{equation}
(k+h^{\vee })\partial _{i}\psi =\left( \sum_{j\neq i=1}^{N+1}\frac{\Omega
_{ij}}{z_{i}-z_{j}}\right) \psi ,\;i=1,...,N+1  \label{n14}
\end{equation}%
Here we denote tensor Casimir
\begin{equation}
\Omega _{ij}=B_{ab}^{-1}(x^{a})_{i}\otimes ^{s}(x^{b})_{j}  \label{n15}
\end{equation}%
(the lower indices $i,j$ indicate the spaces of the tensor product in $V$
where the generators $x^{a}$ act.) and $z_{N+1}=0$. Recall that the vectors $%
u_{0}\in M_{p_{0},k}[0]$ and $u_{N+1}\in M_{p_{N},k}[0]$ have grade 0 . Here
we use the super tensor product which for two matrices $A_{\alpha \gamma }$
and $B_{\beta \delta }$ is defined as $(A\otimes ^{s}B)_{\alpha \beta
}^{\gamma \delta }=(-1)^{\beta (\alpha +\gamma )}A_{\alpha \gamma }B_{\beta
\delta }$, where the indices lifted to exponential of $(-1)$ are parities of
corresponding indices in $%
\mathbb{Z}
_{2}$ graded vector spaces. The main difference compared to the usual non
superalgebras and irreducible finite dimensional highest weight modules is
that $\Omega _{ij}$ can act now non diagonally on the modules. In this sense
they are not eigenvalue numbers but operators. With the assumption that $%
u_{N+1}$ is the vector of scalar representation (at least in the sense
described in the footnote) the last term in (\ref{n141}) disappears, and the
equation we will deal with in what follows
\begin{equation}
(k+h^{\vee })\partial _{i}\psi =\sum_{j\neq i=1}^{N}\frac{\Omega _{ij}}{%
z_{i}-z_{j}}\psi ,\;i=1,...,N  \label{n16}
\end{equation}

The proof of the theorem claiming (\ref{n16}) for correlation functions for
superalgebras with non-semisimple modules is a copy of the proof in the case
of simple modules over usual Lie algebras. The proof uses commutation
relations (\ref{crcur}) and the fact that $u_{0},u_{N+1}$ are zero grade
states.

Looking for solutions for $\psi \in V^{\mathfrak{g}}$ is not the only
option. One can get a set of solutions when $\psi $ is projected onto some
weight subspace $\psi \in V^{\lambda }$ of weight $\lambda $. Usually, when
the spaces $M_{p_{i}}$ are highest weight ones $\mu _{i}$, the solutions
with values in the space $(V^{\mathfrak{n}^{+}})^{\lambda }$ are considered.
If $\lambda =\sum \mu _{i}-\mu $, $\mu =\sum n_{i}\alpha _{i}$, $\alpha
_{i}\in Q^{+}$, the value $|\mu |=\sum n_{i}$ is called level of the equation%
\footnote{%
It will be interesting to find a direct way to obtain non zero level
solution from the zero level solutions ones, as it was done in non-super
case \cite{Var}}. Usually level one solutions for $N=3$ already give
solutions with a basis of hypergeometric functions. But in order to see such
hypergeometric solutions in $V^{\mathfrak{g}}$, one has to take at least $%
N=4 $ correlation functions.

Important particular case of KZ equation when it becomes an ordinary
differential equation, is the $N=3$ case. As one can show (see e.g. \cite%
{EFK}), in this case any solution of KZ equation can be written as
\begin{equation*}
\psi(z_{1},z_{2},z_{3})=(z_{1}-z_{3})^{(\Omega_{12}+\Omega_{13}+\Omega
_{23})/\kappa}f\left( \frac{z_{1}-z_{2}}{z_{1}-z_{3}}\right)
\end{equation*}
where $f(z)\in V$ satisfies the differential equation%
\begin{equation}
\kappa\partial_{z}f=\left( \frac{\Omega_{12}}{z}+\frac{\Omega_{23}}{z-1}%
\right) f  \label{3kz}
\end{equation}

For the irreducible modules $M_{q_{1}},...,M_{q_{N}}$ of highest/lowest
weight there is a classification and explicit form of solutions of KZ
equation for specified level of weights in root lattice grading. Level zero
solution is always of the form%
\begin{align*}
\Psi _{0}(z_{1},...z_{N})& =\psi _{0}(z_{1},...z_{N})v,\;v=\mu _{1}\otimes
\mu _{2}...\otimes \mu _{N}, \\
\psi _{0}(z_{1},...z_{N})& =\prod\limits_{i<j}(z_{i}-z_{j})^{\mu _{i}\mu
_{j}/2\kappa }
\end{align*}
Solutions of higher levels of KZ equations in the case of highest or lowest
weight modules $M_{\lambda _{i}}$ at generic $\kappa $ one can obtain by the
following procedure. (We consider highest weight modules). Define
multi-valued function
\begin{equation*}
\phi _{1}(z_{1},...z_{N},t)=\prod\limits_{i=1}^{N}(t-z_{i})^{\mu _{i}/\kappa
}
\end{equation*}%
and fix a closed contour $C$ in $t$ complex plane not containing any of $%
z_{i}$, and having a continuous branch along $C$. Example of such contour is
Pochhammer contour for two $z_{a},z_{b}.$ Existence and classification of
such contours is known for semisimple case, but is a non trivial question
for non semisimple case. Then a general level one solution $\Psi
_{1}(z_{1},...z_{N})$ can be obtained as
\begin{equation}
\Psi _{1}(z_{1},...z_{N})=\psi _{0}(z_{1},...z_{N})\sum_{r=1}^{N}\left(
\int_{C}dt\phi _{1}(z_{1},...z_{N},t)\frac{1}{t-z_{r}}\right) f_{r}v
\label{integr}
\end{equation}%
where $v=v_{1}\otimes ...\otimes v_{N}$ is the highest weights tensor
product, and the step operator $f_{r}$ acts on the $r$th component of tensor
product. The proof is by direct calculations. Explicit realization of this
solution gives rise to integral representations of hypergeometric functions $%
_{2}F_{1}$. Level $l$ solution can be similarly generated by integration of
operator valued differential $l$-forms. The answer in this case is much more
involved \cite{Var}.

For the case of semisimple categories of finite dimensional $\mathfrak{g}$%
-modules at generic level $\kappa $ the most important statement says that
the monodromy of KZ equations gives rise to braided tensor categories, and
that they equivalent to the categories of specific quantum group
representation. One of the ways to see it for generic level case was worked
out by Schechtman and Varchenko \cite{ScheVar} using the integral formulas
of the KZ solutions by analysis of geometry of integration cycles. Can the
same be done in the case of non-semisimple categories of $\mathfrak{g}$%
-modules when solutions involve logarithms? We are going to address this
question elsewhere.

All the construction above treats $z_{i}$ as formal variables. There is a
theorem proved for KZ equations in semisimple case that $\psi $ is an
analytic function of $z_{i}$ in the region $|z_{1}|>|z_{2}|>...>0$. This
analyticity should be modified in the non semisimple case because of
presence of logarithms in intertwiners mode expansions.

Consistency and $\mathfrak{g}$-invariance of KZ equation, as in semisimple
case, follows from $\mathfrak{g}$-invariance of Casimir operator. It has an
important practical application: in order to find the full set of
independent KZ equations for a given correlation function one should find
the basis of invariants of the space $V$ -- the set of tensor product
vectors annihilated by all the generators of $\mathfrak{g}$, and then
project the equations on these vectors. One can find some examples of such
calculations in Appendix \ref{subCorf}. Explicit construction of tensor
category structures of solutions of KZ equations requires calculations up to
$N=4$ -- four point correlation functions.

The final goal is investigation of monodromy properties of solutions of KZ
equation. By this we mean the following. The system of KZ equations being
consistent can be interpreted as a flat connection in the trivial vector
bundle with the fiber $V$ over the configuration space $X_{N}=%
\{(z_{1},z_{2},...,z_{N})\in
\mathbb{C}
^{N}\;|\;z_{i}\neq z_{j}\}$. For any path $\gamma :[0,1]\rightarrow X_{N}$
we denote by $M_{\gamma }$ the operator of holonomy along $\gamma $. It can
be considered as an operator in $V$ and it depends only on homotopy class of
$\gamma $, or as operator of analytic continuation along $\gamma $. From $%
\mathfrak{g}$-invariance of $\Omega $ follows that for any $\gamma $ $%
M_{\gamma }:V\rightarrow V$ is a $\mathfrak{g}$-homomorphism. If $V$ is
completely reducible, then it means that $M_{\gamma }$ preserves subspace of
singular vectors in $V$ and is uniquely defined by its action on this
subspace.

\section{Drinfeld category of $\mathfrak{gl}(1|1)$ modules}

\label{section3}

In this section we consider the KZ equation as an equation on functions
\begin{equation*}
\psi (z_{1},...,z_{N}):%
\mathbb{C}
^{N}\setminus \{Diag\}\rightarrow V[[\kappa ^{-1}]]
\end{equation*}%
where the set of points $\{Diag\}:z_i=z_j, i\neq j$ are removed from the domain $\mathbb{C}^N$. The functions are valued in $V[[\kappa ^{-1}]]$, where $V=V_{1}\otimes ...\otimes V_{N}$ is a
tensor product of moduli representation spaces $V_{i}$ of the superalgebra $%
\mathfrak{gl}(1|1)$. We define abelian tensor Drinfeld supercategory $%
\mathcal{D}$ of subset of finite dimensional non semisimple $\mathfrak{gl}%
(1|1)$-moduli with all moduli homomorphisms as the category morphisms, and
construct its braided tensor category structure.

The superalgebra $\mathfrak{gl}(1|1)$ is the algebra of endomorphisms of the
vector superspace $%
\mathbb{C}
^{1|1}$ $\mathfrak{gl}(1|1)=span\{E,N,\psi ^{+},\psi ^{-}\}$ with two
dimensional even $\mathfrak{gl}(1|1)_{\overline{0}}=span\{E,N\}$ and two
dimensional odd $\mathfrak{gl}(1|1)_{\overline{1}}=span\{\psi ^{+},\psi
^{-}\}$ subspaces written in the superalgebra basis. The commutation
relations of the algebra, explicit form of the basis in the representations
we consider are collected in the Appendix \ref{subgl11}.

The objects of $\mathcal{D}$ include (isomorphism classes of) three types of
modules - typical $\mathcal{T}_{e,n}$, and atypical $\mathcal{P}_{n},%
\mathcal{A}_{n}$ four- and one-dimensional modules (see the Appendix \ref%
{subgl11} for description of the meaning of moduli parameters $e,n$). We
impose the restrictions on the parameters $e_{i}$ of the set of typical
modules $\mathcal{T}_{e_{i},n_{i}}$ to be such that three conditions are
satisfied. First, $e_{i}\kappa ^{-1}\notin
\mathbb{Z}
$, just by definition of the typical module. Second, $\kappa
^{-1}\sum_{k}e_{i_{k}}\notin
\mathbb{Z}
\backslash \{0\}$ for any subset of the set of typical modules $\mathcal{T}%
_{e_{i},n_{i}}$ in the category (the reason for this restrictions will be
explained below). And the third, for any subset of typical modules $%
\sum_{k}e_{i_{k}}$ is equal either to some $e_{n}$ or to 0. The reason for
the latter restriction is that such choice guarantee the closure of the set
of objects under the tensor product decomposition. This decomposition is
well known (see for example \cite{CRarhy}, eq. (2.11)-(2.16))
\begin{align}
\mathcal{A}_{n}\otimes \mathcal{A}_{n^{\prime }}& =\mathcal{A}_{n+n^{\prime
}},\;\mathcal{A}_{n}\otimes \mathcal{T}_{e,n^{\prime }}=\mathcal{T}%
_{e,n+n^{\prime }}  \label{gl1} \\
\mathcal{T}_{e,n}\otimes \mathcal{T}_{e^{\prime },n^{\prime }}& =\mathcal{T}%
_{e+e^{\prime },n+n^{\prime }+1/2}\oplus \Pi \mathcal{T}_{e+e^{\prime
},n+n^{\prime }-1/2},  \notag \\
\mathcal{T}_{e,n}\otimes \mathcal{T}_{-e,n^{\prime }}& =\mathcal{P}%
_{n+n^{\prime }},\;\mathcal{A}_{n}\otimes \mathcal{P}_{n^{\prime }}=\mathcal{%
P}_{n+n^{\prime }},  \notag \\
\mathcal{T}_{e,n}\otimes \mathcal{P}_{n^{\prime }}& =\Pi \mathcal{T}%
_{e,n+n^{\prime }+1}\oplus 2\mathcal{T}_{e,n+n^{\prime }}\oplus \Pi \mathcal{%
T}_{e,n+n^{\prime }-1},  \notag \\
\mathcal{P}_{n}\otimes \mathcal{P}_{n^{\prime }}& =\mathcal{P}_{n+n^{\prime
}+1}\oplus 2\Pi \mathcal{P}_{n+n^{\prime }}\oplus \mathcal{P}_{n+n^{\prime
}-1}.  \notag
\end{align}%
The functor $\Pi $ for some modules on the right hand side denotes parity
reversion of the $%
\mathbb{Z}
_{2}$ grading of even and odd module subspaces. The obvious requirements on
the set of parameters $n_{i}$ of the modules in the category similarly
follows by closure of the tensor product decomposition. No other
restrictions on the moduli parameters $n_{i}$ are imposed.

Of course there are infinitely many other finite dimensional indecomposable $%
\mathfrak{gl}(1|1)$-modules, but our choice seems to be the minimal set of
(isomorphism classes of) modules closed under the tensor product
decomposition with a non trivial braiding structure described below.

The indecomposable modules $\mathcal{P}_{n}$ are called projective, because
they are projective covers for $\mathcal{A}_{n}$. The typical modules $%
\mathcal{T}_{e,n}$ are their own projective covers. Indecomposable structure
of the modules can be found in the same reference \cite{CRarhy}, Section
2.2. The modules of our category are finitely generated and are semisimple
under the action of the even part of the superalgebra. Some properties of
such categories of $\mathfrak{gl}(m|n)$-modules were reviewed in \cite{Br}.

We see that one should include in the category the modules obtained by the
parity change functor $\Pi $. It means the above tensor rules should be
completed by the copy of them with the obvious action of $\Pi $, which we
omit for brevity. All the statements below will be proved for the part of
tensor ring (\ref{gl1}), and is identical for its parity change analog. The
standard parity for the modules are chosen in the following way. We assume
the highest weight of the two dimensional typical module $\mathcal{T}_{e,n}$
($e\neq 0$) to be grassmann even, as well as the one dimensional atypical
module $\mathcal{A}_{n}$, and the top vector of the projective module $%
\mathcal{P}_{n}$ (see Appendix \ref{subgl11}) to be also even.

The structure of braided tensor category $(\mathcal{D},\times ,\mathbf{1}%
,\lambda ,\rho ,\sigma )$ is defined as follows. The \ bifunctor $\mathcal{%
D\times D\rightarrow D}$ is the tensor product of the modules that was
described above.The unit object of $\mathcal{D}$ is $\mathbf{1}=\mathcal{A}%
_{0}$ is simple, and as follows from (\ref{gl1}) the functorial isomorphisms
$\lambda :\mathbf{1}\otimes U\tilde{\longrightarrow}U$, $\rho :U\otimes
\mathbf{1}\tilde{\longrightarrow}U$ are trivial. Below we will define and
prove the existence of invertible associator - functorial isomorphism $%
\alpha _{X,Y,Z}:(X\otimes Y)\otimes Z\tilde{\longrightarrow}X\otimes
(Y\otimes Z)$ for any triple of objects $X,Y,Z\in Obj(\mathcal{D})$. This
isomorphism is defined using asymptotic solutions of KZ equations. The
braiding $\sigma :X\otimes Y\rightarrow Y\otimes X$ of any two objects is
defined by $\sigma =Pe^{i\pi \Omega _{12}/\kappa }$ where $P$ is graded
permutation. The prove of coherence theorem for associator, i.e. pentagon
and triangle relations for monoidal structure becomes standard after the
explicit construction of associator, as well as the proof of hexagon
relation for braiding.

First we briefly recall the monodromy structure and asymptotic solutions of
KZ equations for semisimple category of modules. We follow and recapitulate
the main steps presented in \cite{Var2}, Section 2. The system of KZ
equations can be interpreted as a flat connection in a trivial vector bundle
with a fiber $V=V_{1}\otimes ...\otimes V_{N}$, $V_{i}$ are objects of $%
\mathcal{D}$, over the configuration space $X_{N}=\{(z_{1},...,z_{N})\in
\mathbb{C}
^{N}\;|\;z_{i}\neq z_{j}\}$. For any path $\gamma :[0,1]\rightarrow X_{N}$
one denotes by $M_{\gamma }:V\rightarrow V$ the operator of holonomy along $%
\gamma $, which can be considered as analytic continuation of KZ equation
solutions $\psi (z_{1},...,z_{N})$ along $\gamma $. $M_{\gamma }$ is $%
\mathfrak{g}$-homomorphism since the tensor Casimir operator $\Omega $ of KZ
equation is $\mathfrak{g}$-invariant. Operator $M_{\gamma }$ with $\gamma
(0)=\gamma (1)=z^{0}=(z_{1}^{0},...,z_{N}^{0})$ is called the monodromy
operator. We have such $M_{\gamma }$ as a monodromy representation of the
fundamental group $\pi _{1}(X_{N},z^{0})$ in $V$. The dependence on the base
point $z^{0}$ can be eliminated by conjugation, because $X_{N}$ is
connected. But the fundamental group $\pi _{1}(X_{N})$ is well known -- it
is $PB_{N}$ -- pure braid group. Moreover, one can construct the
homomorphism of braid group $B_{N}\rightarrow \pi _{1}(X_{N}/S_{N})$ where $%
S_{N}$ is the symmetric group: if we choose the $z^{0}$ such that $%
z_{i}^{0}\in
\mathbb{R}
$ and $z_{1}^{0}>z_{2}^{0}>...>z_{N}^{0}$ then the action of $b_{i}$
generator of $B_{N}$ on $z^{0}$ corresponds to transposition of $z_{i}^{0}$
and $z_{i+1}^{0}$ (say, $z_{i+1}^{0}$ and $z_{i}^{0}$ exchange their
locations such that $z_{i}^{0}$ passes above $z_{i+1}^{0}$). For a fixed
base point $z^{0}$ a loop $\gamma $ in $X_{N}/S_{N}$ can be considered as an
element of $B_{N}$. Then we can lift it to a path in $X_{N}$ defining the
operator $\check{M}_{\gamma }=\sigma M_{\gamma }:V\rightarrow V^{\sigma }$,
where $\sigma \in S_{N}$ is the image of $\gamma $ under the map $%
B_{N}\rightarrow S_{N}$ and $V^{\sigma }=V_{\sigma ^{-1}(1)}\otimes
...\otimes V_{\sigma ^{-1}(N)}$. For example, for the $\gamma $ which
exchanges $z_{i}^{0}$ and $z_{i+1}^{0}$ we will have $\check{M}_{i}^{\pm
}(z^{0})=\check{M}_{\gamma _{i}}^{\pm 1}$. The fact that the operators $%
\check{M}_{i}^{\pm }$ called \textit{half monodromy operators} satisfy the
equations
\begin{eqnarray*}
\check{M}_{i}^{\pm }\check{M}_{i}^{\mp } &=&I, \\
\check{M}_{i}^{\pm }\check{M}_{i+1}^{\pm }\check{M}_{i}^{\pm } &=&\check{M}%
_{i+1}^{\pm }\check{M}_{i}^{\pm }\check{M}_{i+1}^{\pm }
\end{eqnarray*}%
follows from the relation $\gamma _{i}\gamma _{i+1}\gamma _{i}=\gamma
_{i+1}\gamma _{i}\gamma _{i+1}$ in the fundamental group of $X_{N}/S_{N}$.

The (half)monodromy operators being independent on the choice of base point,
can be calculated with a specific choice of it. One of the convenient
choices of the base point is $z^{0}:z_{1}^{0}\gg z_{2}^{0}\gg ...\gg
z_{N}^{0}$. We will need also another choice of the base point for $N=3$
correlation function below. We fix the region $D\subset X_{N}$, $%
D=\{z=(z_{1},...,z_{N})\in
\mathbb{R}
^{N}\;|\;z_{1}>...>z_{N}\}$. There is an isomorphism between the space of $V$%
-valued solutions $\Gamma _{f}(D,V_{KZ})$ of the KZ equation in the region $%
D $ and $V$ : for any $z\in D$ the solution $\psi (z)$ is this isomorphism.
It is useful to make the following change of variables.%
\begin{align}
u_{i}& =\frac{z_{i}-z_{i+1}}{z_{i-1}-z_{i}},\;i=2,...,N-1  \label{35} \\
u_{1}& =z_{1}-z_{2},\;u_{N}=z_{1}+...+z_{N}  \notag
\end{align}%
All $u_{i}$ are positive on $D$. One can see that $(z_{1},...,z_{N})%
\rightarrow (u_{1},...,u_{N})$ is one to one map with inverse polynomial
mapping, therefore any analytic function $f(z)$ on $D$ can be considered as
analytic function of $u$ on some subset $D_{u}\subset
\mathbb{C}
^{N}$ - the image of the mapping, and closure of $D_{u}$ contains the
origin. The change of variables (\ref{35}) is chosen so that if we have a
curve $z(t)$ such that $z(t)\rightarrow 0$ when $t\rightarrow 0$, then the
condition $z_{i}(t)/z_{i+1}(t)\rightarrow \infty $ for $i=1,...,N-1$,
implies $u_{i}(t)\rightarrow 0$ for $i=1,...,N$.

We can define now the limit $\lim_{z_{1}\gg ...\gg z_{N}}f(z)=v$ as a vector
which satisfies $\lim_{u_{i}\rightarrow 0}f(u)=v$ for $i=1,...,N$ with $f$
being written in terms of new variables $u_{i}$.

We define the asymptotic $f\sim \phi _{1}(z)v$ of a smooth vector valued
function $f(z)$ as the $z_{1}\gg ...\gg z_{N}$ limit of $f$ in \ $D$, for
some scalar function $\phi _{1}(z)$ and a vector $v\in V$, if
\begin{equation}
f(z)=\phi _{1}(z)(v+o(z))  \label{36}
\end{equation}%
where $o(z)$ considered as a $V$-valued function of $u$ in some neighborhood
of the origin is regular and $o(u=0)=0$. We will sometimes put $z_{N}=0$. If
$f$ is translation invariant then $\lim_{z_{1}\gg ...\gg
z_{N}}f(z)=\lim_{z_{1}\gg ...\gg 0}f(z)$.

Another region we need is $D_{0}(z):z_{1}-z_{2}\ll z_{2}-z_{3}\ll ...\ll
z_{N-1}-z_{N}$ and as above we define the asymptotic of a function $f(z)$ in
the region $D_{0}(z)$ as $f\sim \phi _{0}(z)v$ if $f(z)=\phi _{0}(z)(v+o(z))$
where $o(z)$ considered as a $V$-valued function of $u$ in some neighborhood
of the point $u_{i}\rightarrow \infty $.

The special case important for the proof of associator existence is $N=3$.
The KZ equation takes the form of ordinary differential equation in one
variable. In terms of the variables (\ref{35}) $u_{1}=z_{1}-z_{2}$, $u_{2}=%
\frac{z_{2}-z_{3}}{z_{1}-z_{2}}$, $u_{3}=z_{1}+z_{2}+z_{3}$ the KZ equations
look like
\begin{eqnarray}
\kappa \partial _{u_{1}}\psi &=&\frac{\Omega _{12}+\Omega _{13}+\Omega _{23}%
}{u_{1}}\psi  \label{KZ31} \\
\kappa \partial _{u_{2}}\psi &=&\left( \frac{\Omega _{12}}{u_{2}+1}+\frac{%
\Omega _{23}}{u_{2}}\right) \psi  \notag \\
\kappa \partial _{u_{3}}\psi &=&0  \notag
\end{eqnarray}%
We introduce the function $f$ defined by\footnote{%
This function is well defined with the choice of the branch of logarithm fixed above because the operators $\Omega _{ij}$ acting in
the space $V$ have nilpotent non diagonalizable part.}
\begin{equation*}
\psi (z_{1},z_{2},z_{3})=(z_{1}-z_{3})^{(\Omega _{12}+\Omega _{13}+\Omega
_{23})/\kappa }f\left( \frac{z_{1}-z_{2}}{z_{1}-z_{3}}\right)
\end{equation*}%
Using the fact that all $\Omega _{ij}$ commute with $\Omega _{12}+\Omega
_{13}+\Omega _{23}$ one can see by direct calculation that $%
f=u_{1}^{-(\Omega _{12}+\Omega _{13}+\Omega _{23})/\kappa }\psi $ depends
only on $x=\frac{1}{u_{2}+1}$ and is $u_{1},u_{3}$ independent. Thus we get
one ODE for the $V$-valued function $f(x)$
\begin{equation}
\kappa \partial _{x}f(x)=\left( \frac{\Omega _{12}}{x}+\frac{\Omega _{23}}{%
x-1}\right) f(x)  \label{KZf}
\end{equation}%
The asymptotic regions $D_{0}(z),D_{1}(z)$ correspond to $x\rightarrow 0$
and $x\rightarrow 1$ respectively. The existence of asymptotic solutions of
KZ equation as they are defined above is the main tool for the proof of
existence of associator.

\begin{theorem}
\label{thm1} Let $V=V_{1}\otimes V_{2}\otimes V_{3}$ where $\{V_{i}\}$ --
any combination from the set $\{\mathcal{A},\mathcal{P},\mathcal{T}\}$. If $%
e_{i}\notin
\mathbb{Z}
$\textbf{\ }and $e_{1}+e_{2}\notin
\mathbb{Z}
\backslash \{0\}$ in the case $V_{i}=\mathcal{T}$ $,i=1,2$\ in $V$, then for
every eigenvector $v\in V$ of $\Omega _{12}$ there exists unique asymptotic
solution of (\ref{KZf}) around 0 corresponding to $v$ and this
correspondence gives isomorphism $\phi _{0}:\Gamma _{f}(D,V_{KZ})\rightarrow
V$.
\end{theorem}

\textit{Proof. }The proof is based on straightforward linear algebra
manipulations which we moved to Appendix A.\textit{\ }We apply Lemma \ref%
{lemma: A1} or \ref{lemma: A2} (see Appendix A), considering all possible 6
combinations (up to a permutation) of $V_{1},V_{2}$: $\mathcal{T}%
_{e_{1},n_{1}}\otimes \mathcal{T}_{e_{2},n_{2}}$, $\mathcal{T}%
_{e_{1},n_{1}}\otimes \mathcal{P}_{n_{2}}$, $\mathcal{P}_{n_{1}}\otimes
\mathcal{P}_{n_{2}}$, $\mathcal{T}_{e_{1},n_{1}}\otimes \mathcal{A}_{n_{2}}$%
, $\mathcal{P}_{e_{1},n_{1}}\otimes \mathcal{A}_{n_{2}}$, $\mathcal{A}%
_{n_{1}}\otimes \mathcal{A}_{n_{2}}$. The explicit form of the function
solution $\phi (x)$ is not important at this point, but one can find it in
the Appendix A. All we have to do is to check, case by case, the
applicability of Lemmas \ref{lemma: A1},\ref{lemma: A2}. Isomorphism to the
space $\Gamma _{f}(D,V_{KZ})$ of KZ solution follows by linearity. The
following data is obtained by direct diagonalization of $\Omega _{12}$ on
the basis of $V_{1}\otimes V_{2}$.

1. $\mathcal{T}_{e_{1},n_{1}}\otimes \mathcal{T}_{e_{2},n_{2}}$.

When $e_{2}+e_{1}\notin
\mathbb{Z}
$ there are no Jordan blocks and the eigenvalues are $\lambda _{1}=\delta
_{12}^{++}$, $\lambda _{2}=\delta _{12}^{--}$, with two eigenvectors for
each of them. Here and below $\delta _{ij}^{\alpha \beta
}=e_{i}e_{j}+e_{i}(n_{j}+\beta /2)+e_{j}(n_{i}+\alpha /2)$. The difference $%
\lambda _{1}-\lambda _{2}=e_{1}+e_{2}\notin
\mathbb{N}
$ and by the Lemma \ref{lemma: A1} there are four asymptotic solutions for
four different eigenvectors.

When $e_{2}+e_{1}=0$ there is one eigenvalue $e_{1}(n_{2}-n_{1})-e_{1}^{2}$
with two eigenvectors without Jordan block and two other ones with Jordan
block of size 2. By the Lemma \ref{lemma: A2} there are four asymptotic
solutions.

We cannot prove existence of asymptotic solutions using Lemma \ref{lemma: A1}
in the case $e_{2}+e_{1}\in
\mathbb{Z}
\backslash \{0\}$, but this case, from the perspective of affine Lie
superalgebra, exactly corresponds to what we call non generic case of
representations \cite{CRbeyond}.

2. $\mathcal{T}_{e_{1},n_{1}}\otimes \mathcal{P}_{n_{2}}$

The set of eigenvalues are $\lambda _{1}=e_{1}(n_{2}-1)$ and $\lambda
_{2}=e_{1}(n_{2}+1)$ with the difference $2e_{1}\notin
\mathbb{N}
$. Each of them \ correspond to two eigenvectors without Jordan block and
one Jordan block of size 2. By the Lemmas \ref{lemma: A1},\ref{lemma: A2}
there are asymptotic solutions for each eigenvector.

3. $\mathcal{P}_{n_{1}}\otimes \mathcal{P}_{n_{2}}$

There is one eigenvalue $\lambda =0$ with the following structure of
eigenvectors: there are 3 Jordan blocks of rank 2, one Jordan block of rank
3 and 7 eigenvectors without Jordan block structure. Again the condition $%
\lambda +%
\mathbb{N}
$ is not an eigenvalue is satisfied, therefore by Lemmas \ref{lemma: A1},\ref%
{lemma: A2} there are asymptotic solutions corresponding to each eigenvector.

4. $\mathcal{T}_{e_{1},n_{1}}\otimes \mathcal{A}_{n_{2}}$

There is one eigenvalue $\lambda =e_{1}n_{2}$ with two different
eigenvectors without a Jordan block. Lemma \ref{lemma: A1} is applicable.

5. $\mathcal{P}_{n_{1}}\otimes \mathcal{A}_{n_{2}}$

There is one eigenvalue $\lambda =0$ with four different eigenvectors
without a Jordan block. Lemma \ref{lemma: A1} is applicable.

6. $\mathcal{A}_{n_{1}}\otimes \mathcal{A}_{n_{2}}$

There is one eigenvalue $\lambda =0$ with one eigenvector. Lemma \ref{lemma:
A1} is applicable.

$\square $

\begin{theorem}
\label{thm2} The same claim as in the Theorem \ref{thm1}, with the same
restrictions on the parameters of typical modules $\mathcal{T}$ appearing as
$V_{i},i=2,3$ in $V$, is valid for existence and uniqueness of asymptotic
solutions of KZ equation (\ref{KZf}) around $x=1$.
\end{theorem}

\textit{Proof.} The proof is based on the Lemma \ref{lemma: A3} (see
Appendix A) that replaces the Lemmas \ref{lemma: A1},\ref{lemma: A2} in the
proof of Theorem \ref{thm1}.

$\square $

As we see, there are specific cases $2e_{i}\in
\mathbb{Z}
$ and $e_{1}+e_{2}\in
\mathbb{Z}
\backslash \{0\}$ for parameters of typical representations when we are not
able to guarantee the existence and uniqueness of asymptotic solutions by
Lemmas \ref{lemma: A1},\ref{lemma: A2},\ref{lemma: A3}. We notice that for
affine $\widehat{\mathfrak{gl}}(1|1)$ (where the we always can put $\kappa
=k=1$) these cases correspond to reducibility of the induced affine modules,
and as we said above, we exclude these cases in the process of derivation of
KZ equation.

\begin{proposition}
With the restrictions on the parameters of typical modules as in the Theorem %
\ref{thm1} there is an isomorphisms of the spaces
\begin{equation}
\alpha _{1,2,3}:(V_{1}\otimes V_{2})\otimes V_{3}\tilde{\longrightarrow}%
\Gamma _{f}(D,V_{KZ})\tilde{\longrightarrow}V_{1}\otimes (V_{2}\otimes V_{3})
\label{asoc}
\end{equation}%
which will serve the associator in the Drinfeld tensor category.
\end{proposition}

\textit{Proof. }The first isomorphism $\phi _{0}:(V_{1}\otimes V_{2})\otimes
V_{3}\tilde{\longrightarrow}\Gamma _{f}(D,V_{KZ})$ is defined by the
correspondence between the eigenvectors of $\Omega _{12}$ in $V$ and
asymptotic solutions of KZ equation (\ref{KZf}) around $x=0$ established by
the Theorem \ref{thm1}. The second isomorphism $\phi _{1}^{-1}:\Gamma
_{f}(D,V_{KZ})\tilde{\longrightarrow}V_{1}\otimes (V_{2}\otimes V_{3})$ is
the inverse of the isomorphism $\phi _{1}$ established by the Theorem \ref%
{thm2}.

$\square $

\textbf{Remark 1. }One can easily see that the associator (\ref{asoc}) is
trivial (equal to 1) when one of the spaces $V_{i},i=1,2,3$ is one
dimensional, as for example in the cases 4,5,6 of the proof of the Theorem %
\ref{thm1}.

\begin{theorem}
\label{thm3} For any quadruple of objects $V_{i},i=1,...,4$ in the $%
\mathfrak{gl}(1|1)$ Drinfeld category $\mathcal{D}$, with the restrictions
on the parameters of typical modules $e_i\notin\mathbb{Z}$, $e_{i}+e_{j}\notin
\mathbb{Z}
\backslash \{0\}$ for any pair $\mathcal{T}_{e_{i},n_{i}},\mathcal{T}%
_{e_{j},n_{j}}$, the isomorphism $\alpha _{1,2,3}$ (\ref{asoc}) satisfies
pentagon equation $((V_{1}\otimes V_{2})\otimes V_{3})\otimes
V_{4}\longrightarrow V_{1}\otimes (V_{2}(\otimes V_{3}\otimes V_{4}))$%
\begin{equation}
\alpha _{id_{1}\otimes 2,3,4}\circ \alpha _{1,2\otimes 3,4}\circ \alpha
_{1,2,3\otimes Id_{4}}=\alpha _{1,2,3\otimes 4}\circ \alpha _{1\otimes 2,3,4}
\label{pen}
\end{equation}
\end{theorem}

The proof is based on decomposition of pentagon diagram into triangle ones,
and each triangle is a commutative diagram which includes as a part the
isomorphism (\ref{asoc}). The proof uses only the fact of existence and
uniqueness of invertible associator irrespectively of details of its
construction from asymptotic solutions. We refer to the books \cite{BK},
p.25, or \cite{ChP}, p.545 for details of the proof, which is independent on
concrete form of asymptotic solutions but only on the fact of their
existence.

$\square $

Recall the standard derivation of braiding $\sigma _{X,Y}$ from
half-monodromy of KZ solutions (See \cite{ChP} Section 16.2 and original
references therein.) Since the solution of KZ equations for $N=2$ is a
function of difference $z_{2}-z_{1}$, one can represent the braid group $%
B_{2}$ generator $\sigma _{12}$ which swaps $z_{1}$ and $z_{2}$, $%
z_{1},z_{2}\in D\subset
\mathbb{C}
^{2}$ by the loop contour $\overrightarrow{z}(s)=(z_{1}(s),z_{2}(s))$, $%
z_{1,2}(s)=a+be^{i\pi s}$, $a=(z_{1}+z_{2})/2$, $b=(z_{1}-z_{2})/2$,
parametrized by $s\in \lbrack 0,1]$. It satisfies $\overrightarrow{z}%
(0)=z_{1}$, $\overrightarrow{z}(1)=z_{2}$. A pull back of the KZ $N=2$
equation written for a one form $dw$ along this contour leads to the
equation
\begin{equation}
\frac{dw}{ds}=\frac{\Omega _{12}}{\kappa }w(s)  \label{br1}
\end{equation}%
with the solution
\begin{equation}
w(s)=e^{\frac{\Omega _{12}}{\kappa }s}w(0)  \label{brsol}
\end{equation}%
As before the exponent is understood here as classical series $\sum \left(
\frac{\Omega _{12}}{\kappa }s\right) ^{n}\frac{1}{n!}$, which converges on $%
Aut(V_{1}\otimes V_{2})$ because of the nilpotency of non diagonal part of $%
\Omega _{12}$ acting on any tensor product of vectors. Therefore if we put $%
s=1$ in the last equation we get the monodromy representation of braid group
\begin{equation}
\rho _{N=2}(\sigma _{12})(v_{1}\otimes v_{2})=Pe^{\frac{\Omega _{12}}{\kappa
}}(v_{1}\otimes v_{2})  \label{rosol}
\end{equation}%
It is straight forward now to generalize this representation of braiding
through half-monodromy of KZ solution to $N>2$.
\begin{equation}
\rho _{N}(\sigma _{i,i+1})(v_{1}\otimes ...\otimes v_{N})=P_{i,i+1}e^{\frac{%
\Omega _{i,i+1}}{\kappa }}(v_{1}\otimes ...\otimes v_{N})  \label{rosol2}
\end{equation}

\begin{theorem}
\label{thm4} For any triple of objects $V_{1},V_{2},V_{3}$ in the Drinfeld
category $\mathcal{D}$ with the restrictions on parameters of $V_{i}=$ $%
\mathcal{T}_{e_{i},n_{i}}$ as above, associator $\alpha _{1,2,3}$ and
braiding $\sigma _{1,2}:V_{i}\otimes V_{j}\longrightarrow V_{j}\otimes V_{i}$%
, $\sigma _{1,2}=P\exp (i\pi \Omega _{12}/\kappa )$ where $P$ is super
permutation of spaces, satisfy the hexagon relation $(V_{1}\otimes
V_{2})\otimes V_{3}\longrightarrow V_{2}\otimes (V_{3}\otimes V_{1})$%
\begin{equation}
\alpha _{2,3,1}\circ \sigma _{1,2\otimes 3}^{\pm 1}\circ \alpha
_{1,2,3}=(Id_{2}\otimes \sigma _{1,3}^{\pm 1})\circ \alpha _{2,1,3}\circ
(\sigma _{1,2}^{\pm 1}\otimes Id_{3})  \label{hex}
\end{equation}%
Moreover the half monodromy operators $\check{M}_{1}$ acting on $%
V_{1}\otimes (V_{2}\otimes V_{3})$ defined above coincide with $\alpha
_{1,2,3}^{-1}\sigma _{12}\alpha _{1,2,3}$.
\end{theorem}

The existence of the universal form of the representation of braiding (\ref%
{rosol}), (\ref{rosol2}) allows to apply the same proof as in the case of
semisimple categories. We refer to \cite{ChP}, p.547 for details of the
proof, which follows \cite{Dr0}, \cite{Ko}.

There is an interesting explicit representation of the associator written in
terms of P-exponential. It was suggested by Drinfeld and a proof that this
is indeed an associator can be found in \cite{Saf}%
\begin{equation}
\alpha _{1,2,3}=\lim_{t\rightarrow 0}\left[ t^{-\Omega _{23}/\kappa }P\exp
\left( \frac{1}{\kappa }\int\limits_{t}^{1-t}\left( \frac{\Omega _{12}}{z}+%
\frac{\Omega _{23}}{z-1}\right) dz\right) t^{\Omega _{12}/\kappa }\right]
\label{asocPexp}
\end{equation}%
Unfortunately even in the case of $\mathfrak{gl}(1|1)$ superalgebra an
explicit calculation of this expression is hard and leads to a complicated
series and interesting algebraic structure \cite{Fur} which we will not
discuss here.

Braided tensor structure of this category is standard for modules category
of quasitriangular Hopf algebra: trivial unit object, trivial associator and
unit morphisms, and braiding morphisms $\sigma _{V,W}=PR_{V,W}$ where $P$ is
super permutation. The proof is standard, and doesn't refer to any
particular data and we refer to textbooks, for example to \cite{ChP}. For
the correspondence with the Drinfeld category we mention the functorial
isomorphism $\beta _{X,Y,Z}^{\pm }:X\otimes (Y\otimes Z)\rightarrow Y\otimes
(X\otimes Z)$ defined by
\begin{equation}
\beta _{X,Y,Z}^{\pm }=\alpha (\sigma _{XY}^{\pm 1}\otimes Id_{Z})\alpha ^{-1}
\label{qmonodr}
\end{equation}%
It satisfies
\begin{equation}
\beta _{X,Y,Z}^{\pm }\beta _{Y,X,Z}^{\mp }=Id  \label{qmon1}
\end{equation}%
Then the functorial isomorphisms
\begin{eqnarray}
\beta _{12}^{\pm } &=&\beta _{X,Y,Z\otimes U}^{\pm }:X\otimes (Y\otimes
(Z\otimes U))\rightarrow Y\otimes (X\otimes (Z\otimes U)),  \label{qmon2} \\
\beta _{23}^{\pm } &=&Id_{X}\otimes \beta _{Y,Z,U}^{\pm }:X\otimes (Y\otimes
(Z\otimes U))\rightarrow X\otimes (Z\otimes (Y\otimes U))  \notag
\end{eqnarray}%
satisfy the relation
\begin{equation}
\beta _{12}^{\pm }\beta _{23}^{\pm }\beta _{12}^{\pm }=\beta _{23}^{\pm
}\beta _{12}^{\pm }\beta _{23}^{\pm }  \label{qmon3}
\end{equation}

We can summarise the construction of Drinfeld category by the following
proposition based on the Theorems \ref{thm1},\ref{thm2},\ref{thm3},\ref{thm4}%
.

\begin{proposition}
\label{prop1} The category $\mathcal{D}$ of typical, atypical and projective
$\mathfrak{gl}(1|1)$-modules with the restrictions on typicals with  $e_i\notin\mathbb{Z}$, $(e_{i}+e_{j})/\kappa \notin
\mathbb{Z}
\backslash \{0\}$ is braided tensor category with the structures as
described above.
\end{proposition}

With these structures category $\mathcal{D}$ of $\mathfrak{gl}(1|1)$-modules
will be considered as category of modules of the algebra denoted by $%
\mathcal{A}_{\mathfrak{g},\Omega }$, ($\mathfrak{g=gl}(1|1)$).

\section{Category $\mathcal{C}_{\protect\kappa }$ of $U_{h}(\mathfrak{gl}%
(1|1))$-modules}

\label{section4}

We denote $i\pi \kappa ^{-1}=h$. The structure of quasitriangular $h$-adic
Hopf superalgebra $A=U_{h}(\mathfrak{gl}(1|1))$, $\kappa \in
\mathbb{R}
^{\times }$, is defined by the following commutation relations of its
generators $\psi ^{\pm },N,E$
\begin{equation*}
\{\psi ^{+},\psi ^{-}\}=2\sinh (hE)
\end{equation*}%
\begin{equation}
\lbrack N,\psi ^{\pm }]=\pm \psi ^{\pm },\;(\psi ^{+})^{2}=(\psi
^{-})^{2}=0,\;[E,X]=0\;\forall X\in U_{h}(\mathfrak{gl}(1|1))  \notag
\end{equation}%
where $\exp (\pm Eh)$ is understood as its Taylor series around $h=0$ ($%
\kappa =\infty $). The Hopf algebra structure is defined as follows.
Coproduct%
\begin{eqnarray}
\overline{\Delta }(E) &=&E\otimes I+I\otimes E,\;\overline{\Delta }%
(N)=N\otimes I+I\otimes N,  \label{qcop} \\
\overline{\Delta }(\psi ^{+}) &=&\psi ^{+}\otimes e^{Eh/2}+e^{-Eh/2}\otimes
\psi ^{+},\;\overline{\Delta }(\psi ^{-})=\psi ^{-}\otimes
e^{Eh/2}+e^{-Eh/2}\otimes \psi ^{-},  \notag
\end{eqnarray}%
counit%
\begin{equation}
\epsilon (E)=\epsilon (N)=\epsilon (\psi ^{\pm })=0,  \label{qco1}
\end{equation}%
and antipode%
\begin{eqnarray}
\gamma (E) &=&-E,\;\gamma (N)=-N,  \label{qanti} \\
\gamma (\psi ^{+}) &=&-e^{Eh/2}\psi ^{+},\;\gamma (\psi ^{-})=-\psi
^{-}e^{-Eh/2},  \notag
\end{eqnarray}%
The algebra $U_{h}(\mathfrak{gl}(1|1))$ is quasitriangular. One can choose
the universal R-matrix $\overline{R}:A\otimes A\rightarrow A\otimes A$ in
the form%
\begin{equation}
\overline{R}=\exp [h(E\otimes E+E\otimes N+N\otimes E)](1-e^{Eh/2}\psi
^{+}\otimes e^{-Eh/2}\psi ^{-})  \label{qRuni}
\end{equation}%
It satisfies the standard quasitriangular Hopf algebra relations%
\begin{eqnarray}
\overline{R}\overline{\Delta }(X) &=&\overline{\Delta }^{op}(X)\overline{R}%
,\;\forall X\in A  \label{qRrel} \\
(\overline{\Delta }\otimes Id)\overline{R} &=&\overline{R}_{13}\overline{R}%
_{23},  \notag \\
(Id\otimes \overline{\Delta })R &=&\overline{R}_{13}\overline{R}_{12},
\notag
\end{eqnarray}

As any quasitriangular Hopf superalgebra $U_{h}(\mathfrak{gl}(1|1))$ induces
braided tensor category structure on the category of finite dimensional
modules provided the latter is closed under the tensor product functor.

\begin{proposition}
Restrictions on $\kappa $ and parameters $e$ of typical modules  $e_i\notin\mathbb{Z}$, $e_{i}+e_{j}\notin
\mathbb{Z}
\backslash \{0\}$ are enough for the category $\mathcal{C}_{\kappa }$ of
(equivalence classes of) the modules $\mathcal{T}_{e,n}^{\kappa },\mathcal{P}%
_{n}^{\kappa },\mathcal{A}_{n}^{\kappa }$ to form a tensor product ring
isomorphic to the tensor product ring (\ref{gl1}) of the modules $\mathcal{T}%
_{e,n},\mathcal{P}_{n},\mathcal{A}_{n}$. (See Appendix B \ref{appB} for
definition of the tensor category $\mathcal{C}_{\kappa }$ in a specified
basis.)
\end{proposition}

We check this by direct calculation in Appendix B \ref{appB} using explicit
basis of three types of representations. It is shown that with the
restrictions on parameters mentioned in the theorem the same tensor product
decomposition works in the quantum case, and the tensor rings are isomorphic.

\section{Proof of braided tensor equivalence}

\label{section5}

The main result of this paper is the following theorem.

\begin{theorem}
\label{thm7} The categories of modules $\mathcal{D}$ and $\mathcal{C}%
_{\kappa }$ with the restrictions on objects of typical modules $e_i/\kappa\notin\mathbb{Z}$, $e_i/\kappa + e_j/\kappa\notin\mathbb{Z}\backslash \{0\}$ are braided tensor equivalent categories.
\end{theorem}

Since our proof of this theorem follows \cite{Gee} , we have change the
approach to KZ equation to a more general one used in \cite{Gee}. In stead of
the KZ equation (\ref{n16}) for correlation functions $\psi $ built on
intertwiners of $\mathfrak{g}$-modules consider the equation - \ we will
call it KZ$_{\mathfrak{g}}$- one can consider KZ equation for superalgebra
valued element $\omega \in (U(\mathfrak{g}))^{\otimes N}[[h]]$ of the form

\begin{equation}
\frac{1}{h}\partial _{i}\omega =\sum_{j\neq i=1}^{N}\frac{\Omega _{ij}}{%
z_{i}-z_{j}}\omega ,\;i=1,...,N  \label{n16a}
\end{equation}
This gives rise to the topologically free quasitriangular quasi-Hopf
superalgebra $\mathcal{A}_{\mathfrak{g},\Omega }$ with the braiding defined
as $e^{h\Omega }$ and the coassociator defined by the monodromy of solutions
of the equation (\ref{n16a}). \ We refer to the standard description of this
algebra in \cite{Dr1} for non-super case, and to it straight forward
generalization for the super case \cite{Gee} \ Section 4. The Drinfeld
category $\mathcal{D}$ is a category of topologically free modules over $%
\mathcal{A}_{\mathfrak{g},\Omega }$.

The equivalence partner for the algebra $\mathcal{A}_{\mathfrak{g},\Omega }$
is the Drinfeld-Jimbo $h$-adic quantum superalgebra $U_{h}(\mathfrak{g})$.
Its structure in our specific case was described in the previous section.

The theorem proved in \cite{Dr2}, which can be modified to the
superalgebra case at hand, claims that if two topological algebras $%
\mathcal{A}_{\mathfrak{g},\Omega }$, $U_{h}(\mathfrak{g})$ are gauge
equivalent (we will explain what it means below), then the categories of
their topologically free modules of finite rank are braided tensor
equivalent. Therefore it is enough for us to show gauge equivalence of the
two superalgebras.

The plan of this section is the following. We start with recalling the
standard proof of the gauge equivalence in the case of simple Lie algebras
which one can find in Drinfeld's paper \cite{Dr3} and explain why it is in
general not applicable in the case of superalgebras. After that we explain
the details of Geer's proof \cite{Gee} of gauge equivalence which avoids the
points of Drinfeld's proof problematical for superalgebras, but applicable
for superalgebras of types $A-G$. At the end we argue why a proof found by
Geer for classical superalgebras of types $A-G$ works also for $\mathfrak{gl}%
(1|1)$ case.

\textbf{The Drinfeld's proof.}

We recall a proof of braided tensor
equivalence of $U_{ih}(\mathfrak{g})$ and $\mathcal{A}_{\mathfrak{g},\Omega
} $ for $\mathfrak{g}$ -- \textit{non-super} Lie algebra \cite{Dr3}, (see
also the Section 16 of \cite{ChP}). This proof is based on the proof of
existence of the invertible element $\mathcal{F}_{h}\in (U(\mathfrak{g}%
)\otimes U(\mathfrak{g}))[[h]]$ which implements the twist of the structures
of the algebra $U(\mathfrak{g})[[h]]$ to the structures of $\mathcal{A}_{%
\mathfrak{g},\Omega }$. The algebra $U_{ih}(\mathfrak{g})$ is isomorphic
\textit{as }$%
\mathbb{C}
\lbrack \lbrack h]]$ \textit{algebra }to $U(\mathfrak{g})[[h]]$. First, one
obtains the algebra $(U(\mathfrak{g})\otimes U(\mathfrak{g}))[[h]]$ from $U(%
\mathfrak{g})[[h]]$ by application of the composite homomorphism
\begin{equation*}
\widetilde{\Delta }_{h}:U(\mathfrak{g})[[h]]\tilde{\rightarrow}U_{ih}(%
\mathfrak{g})\rightarrow \overline{\Delta }\rightarrow U_{ih}(\mathfrak{g}%
)\otimes U_{ih}(\mathfrak{g})\tilde{\rightarrow}(U(\mathfrak{g})\otimes U(%
\mathfrak{g}))[[h]]
\end{equation*}%
If one requires that $\widetilde{\Delta }_{h}=\Delta (\bmod h)$ where $%
\Delta $ is the usual comultiplication in $U(\mathfrak{g})$, then using the
fact that $H^{1}(\mathfrak{g},U(\mathfrak{g})\otimes U(\mathfrak{g}))=0$ for
simple Lie algebras, one gets that there must exist $\mathcal{F}_{h}\in (U(%
\mathfrak{g})\otimes U(\mathfrak{g}))[[h]]$ such that
\begin{equation*}
\mathcal{F}_{h}\equiv 1\otimes 1(\bmod h)
\end{equation*}%
and
\begin{equation}
\mathcal{F}_{h}^{-1}\overline{\Delta }(x)\mathcal{F}_{h}=\widetilde{\Delta }%
_{h}(x),\text{ }\forall x\in U(\mathfrak{g})  \label{tw0}
\end{equation}%
Let the image of the universal R-matrix $\overline{R}$ of $U_{ih}(\mathfrak{g%
})\approxeq U(\mathfrak{g})[[h]]$ in $(U(\mathfrak{g})\otimes U(\mathfrak{g}%
))[[h]]$ under $\widetilde{\Delta }_{h}$ be $\widetilde{R}$. The
quasitriangular Hopf algebra $U(\mathfrak{g})[[h]]$ with trivial
coassociator, the coproduct $\widetilde{\Delta }_{h}$ and the R-matrix $%
\widetilde{R}$ can now be twisted by the element $\mathcal{F}_{h}$, giving
quasitriangular quasi-Hopf algebra $U(\mathfrak{g})[[h]]$ with different
comultiplication, different R-matrix and non trivial coassociator. We would
like them to be the same as of the algebra $\mathcal{A}_{\mathfrak{g},\Omega
}$, i.e $\Delta $ - the trivial coproduct of $U(\mathfrak{g})$. The standard
properties of quasitriangular quasi-Hopf algebras are used to prove that all
three structures can fit to the required ones of $\mathcal{A}_{\mathfrak{g}%
,\Omega }$ using the existing twist element $\mathcal{F}_{h}$. This element
implements what we called above the gauge equivalence. Explicitly the twist
equations are%
\begin{equation}
(\epsilon \otimes id)\mathcal{F}_{h}\mathcal{=}(id\otimes \epsilon )\mathcal{%
F}_{h}=1  \label{tw00}
\end{equation}%
\begin{equation}
\mathcal{F}_{h}^{-1}\overline{\Delta }(x)\mathcal{F}_{h}=\Delta (x)
\label{tw1}
\end{equation}%
\begin{equation}
(\mathcal{F}_{h})_{21}^{-1}\overline{R}_{12}(\mathcal{F}_{h})_{12}=R_{12},
\label{tw2}
\end{equation}%
\begin{equation}
(\mathcal{F}_{h})_{23}(1\otimes \Delta )(\mathcal{F}_{h}).\alpha .[(\mathcal{%
F}_{h})_{12}(\Delta \otimes 1)(\mathcal{F}_{h})]^{-1}=1\otimes 1\otimes 1
\label{tw3}
\end{equation}

Therefore the braided equivalence prove is reduced to a proof of existence
of invertible $\mathcal{F}_{h}$ which satisfies the equations (\ref{tw1}) - (%
\ref{tw3}). The equation (\ref{tw3}) is the most important one. However
explicit solution of the equations (\ref{tw1}) - (\ref{tw3}) is a very hard
problem, which requires an explicit form of associator. All we are able to
do in this context is to prove its existence, in a way described above. One
of the problems to repeat these arguments of twist $\mathcal{F}_{h}$
existence for a superalgebra case, is that the vanishing of the first
cohomology $H^{1}(\mathfrak{g},U(\mathfrak{g})\otimes U(\mathfrak{g}))=0$
used above doesn't not hold in general for superalgebras, in particular for $%
\mathfrak{g}=\mathfrak{gl}(1|1)$, (see for example \cite{Moe}). One should
look for a way which avoids the cohomology vanishing arguments. One of such
ways was suggested by N. Geer  \cite{Gee} by superalgebra modification of
quantization procedure suggested by Etingof and Kazhdan (EK) \cite{EK1} -
\cite{EK3}\footnote{%
The complete list of relevant sequel of their papers is longer, but the
others will not be used in our discussion below.}.

\textbf{The Geer's proof.}

We start from an concise outline and the main steps of the proof in \cite%
{Gee} and then provide some details. The EK construction \cite{EK1} includes
two algebras -- the algebra $H=U_{h}(D(\mathfrak{g}))$, the quantisation of
quantum double, and the algebra $\overline{U}_{h}(\mathfrak{g})$. The latter
is a quantum double which in general admits a non trivial $h$-adic topology%
\footnote{%
In \ \cite{EK1} the $\overline{U}_{h}(\mathfrak{g})$ quantisation was built
to be applied to infinite dimensional Lie algebras, when the quantization $H$
doesn't cannot be applied, but both quantizations work for finite
dimensional Lie algebras.}. Both quantizations are generalized in the
Sections 5,6,7 for the types $A-G$ superalgebras in \cite{Gee} and shown to
be equivalent. The main features which make this way of quantization
effective is \textit{commutativity with quantum double}, by construction, of
the first quantisation, and \textit{functoriality} of the second (Section 8
of \cite{Gee}).

The next step of the proof (\cite{Gee}, Section 9) is isomorphism of the two
equivalent EK quantizations $U_{h}(\mathfrak{g})$ to the standard
Drinfeld-Jimbo quantization $U_{h}^{DJ}(\mathfrak{g})$. The proof follows
\cite{EK3} where the assertion was proved for non-super Kac-Moody algebras
case but works for finite Lie algebras as well. The superalgebra case
requires explicit check of additional Serre relations typical for the most of
the quantum superalgebras of types $A-G$.

The final steps of the proof are in the Section 10. (All the references
below are to the sections and equations of the paper \cite{Gee}.) If there
is a gauge isomorphism $\alpha $ between two quasitriangular
quasi-superbialgebras then it induces braided tensor equivalence between
their modules (Theorem 47). Using the previous results on quantization of
double with explicit form of the twist (eq. (32)) one leads to the
conclusion that $U_{h}(\mathfrak{g})$ is a gauge twist $(\mathcal{A}_{%
\mathfrak{g},\Omega })_{\mathcal{F}}$ of $\mathcal{A}_{\mathfrak{g},\Omega }$%
. The collection of these assertions finally leads to the required
conclusion that the categories of topologically free $\mathcal{A}_{\mathfrak{%
g},\Omega }$ and $U_{h}(\mathfrak{g})$ finite dimensional modules are
braided tensor equivalent (Theorem 48).

Now we explain some details of the proof steps described above and point out
specific features of these steps in our $\mathfrak{gl}(1|1)$ case at the end.

The superalgebra $\mathcal{A}_{\Omega ,\kappa }$ is topologically free
quasitriangular quasi-Hopf superalgebra built from $\mathfrak{g}$, and the
Drinfeld category of modules $\mathcal{D}_{\mathfrak{g}}$ is braided tensor
category of its modules with the structures based on the KZ equation, as
described above.

Let $\mathfrak{g}_{+}$ be a finite dimensional superbialgebra and $\mathfrak{%
g}=D(\mathfrak{g}_{+})$ be its double. In Section 5, following \cite{EK1},
Verma modules $M_{\pm }=U(\mathfrak{g})\otimes _{U(\mathfrak{g}_{\pm
})}c_{\pm }$ over $\mathfrak{g}$ are used in order to construct forgetful
functor $F$ from the Drinfeld category $\mathcal{D}_{\mathfrak{g}}$ to the
category of topologically free $%
\mathbb{C}
\lbrack \lbrack h]]$-modules $\mathcal{A}$:%
\begin{equation}
F(V)=Hom_{\mathcal{D}_{\mathfrak{g}}}(M_{+}\otimes M_{-},V)  \label{Fn}
\end{equation}%
The Theorem 12 asserts that it is a tensor functor. More precisely, there
exists a family of isomorphisms $\mathcal{J}_{V,W}$, $V,W\in \mathcal{D}_{%
\mathfrak{g}}$ such that%
\begin{equation}
\mathcal{J}_{U\otimes V,W}\circ (\mathcal{J}_{U,V}\otimes 1)=\mathcal{J}%
_{U,V\otimes W}\circ (1\otimes \mathcal{J}_{V,W})  \label{FcPr}
\end{equation}%
namely
\begin{equation}
\mathcal{J}_{V,W}(v\otimes w)=(v\otimes w)\circ \alpha
_{1,2,34}^{-1}(1\otimes \alpha _{2,3,4})\circ \beta _{23}\circ (1\otimes
\alpha _{2,3,4}^{-1})\circ \alpha _{1,2,34}\circ (i_{+}\otimes i_{-})
\label{FcDef}
\end{equation}%
Here $i_{\pm }$ is a coproduct defined on the highest (lowest) weights of
the Verma modules as $i_{\pm }(v_{\pm })=v_{\pm }\otimes v_{\pm }$ and $%
\beta $ is the morphism given by $\tau e^{\Omega \kappa /2}$. Theorem 12
with a proof copied from \cite{EK1} asserts that $\mathcal{J}_{V,W}$
together with $F$ is a tensor functor. The functor $F$ can be thought of as
a forgetful functor $F(V):V\rightarrow Hom_{\mathcal{D}_{\mathfrak{g}}}(U(%
\mathfrak{g}),V)$. Being a tensor functor it induces a bialgebra structure
on the target. Moreover, it induces superbialgebra structure on $U(\mathfrak{%
g})[[h]]$ and give rise to a Hopf algebra $H$ with structure isomorphic to a
twist $\mathcal{F}\in U(\mathfrak{g})^{\otimes 2}[[h]]$ determined by $%
\mathcal{J}_{V,W}$ (eq. 32) of the usual structure of $U(\mathfrak{g})[[h]]$%
. Its  R-matrix $R=(\mathcal{F}^{op})^{-1}e^{\kappa \Omega /2}\mathcal{F}$.
This $R$ is polarized, i.e. $R\in U_{h}(\mathfrak{g}_{+})\otimes U_{h}(%
\mathfrak{g}_{-})$. The final assertion of this part (Theorem 17) is that $H$
is a quantization of superbialgebra $\mathfrak{g}$. Two important features
of this construction is that $U_{h}(\mathfrak{g}_{\pm })$ are closed under
coproduct, and that this quantization commutes with taking the double: $%
D(U_{h}(\mathfrak{g}_{+}))\cong U_{h}(\mathfrak{g}_{+})\otimes U_{h}(%
\mathfrak{g}_{-})=H$ (Corollary 23). \ We refer to the Section 5 of \cite%
{Gee} for details of this part of the proof steps.

The construction of this first EK quantisation can be preserved in our $%
\mathfrak{gl}(1|1)$ case. The Verma modules in our notations isomorphic to
the typical modules $\mathcal{T}_{e,n}$. The atypical modules are the
quotients of $\mathcal{T}_{0,n}$, and the projectives $\mathcal{P}_{n}$ can
be identified in this construction with $M_{+}\otimes M_{-}\cong \mathcal{T}%
_{e,n_{1}}\otimes \mathcal{T}_{-e,n_{2}}$ , \ $e\neq 0$, $n_{1}+n_{2}=n$.

The second EK quantization is in a sense "dual" to the first EK
quantization. For a finite dimensional superbialgebra with a discrete
topology (given by inverse limit of finite dimensional topological
superspaces) its modules are topological superspaces. For such topological
modules of Drinfeld double one defines the dual Drinfeld category $\mathcal{D%
}_{\mathfrak{g}}^{t}$ with these \ modules as objects and morphisms $Hom_{%
\mathcal{D}_{\mathfrak{g}}^{t}}(V,W)=$ $Hom_{\mathfrak{g}}(V,W)[[h]]$. \ One
can also dualize $i_{\pm }^{\ast }$  the maps $i_{\pm }$. Similarly functor $%
\overline{F}:\mathcal{D}_{\mathfrak{g}}^{t}\rightarrow \mathcal{A}^{t}$ from
dual Drinfeld category to a symmetric tensor category $\mathcal{A}^{t}$ of $%
\mathbb{C}
\lbrack \lbrack h]]$ modules with continuous maps as morphisms, can be
defined now by $\overline{F}(V)=Hom_{\mathcal{D}_{\mathfrak{g}%
}}(M_{-},M_{+}^{\ast }\otimes V)$. \ Similarly to the first EK quantization
the Theorem 26 asserts that together with the isomorphism
\begin{equation}
\overline{\mathcal{J}}_{V,W}(v\otimes w)=(i_{+}^{\ast }\otimes 1\otimes
1)\circ \alpha _{1,2,34}^{-1}(1\otimes \alpha _{2,3,4})\circ \beta
_{23}^{-1}\circ (1\otimes \alpha _{2,3,4}^{-1})\circ \alpha _{1,2,34}\circ
(v\otimes w)\circ i_{-}  \label{FcDefd}
\end{equation}%
it defines a tensor structure on $\overline{F}$. Further steps of the second
EK quantization are parallel to the first one, similarly leading to the Hopf
algebra $\overline{H}$ which is a quantization of $\mathfrak{g}$ (Theorem
28). Using Proposition 9.7 of \cite{EK1} it is proved that there is an
isomorphism of Hopf superbialgebras $H$ and $\overline{H}$

The following theorems summarize the previous constructions of this step.

Theorem 33: There exists a functor from the category of finite dimensional
superbialgebra $\mathfrak{g}$ over $%
\mathbb{C}
$ and the category of quantum universal enveloping superalgebra over $%
\mathbb{C}
\lbrack \lbrack h]]$ such that $\mathfrak{g}$ is mapped to $U_{h}(\mathfrak{g%
})$ which is the second EK quantization.

Theorem 34: There exist a functor from the category of quasitriangular
finite dimensional superbialgebra $(\mathfrak{g},r)$ over $%
\mathbb{C}
$, where $r$ is classical R-matrix, to the category of quasitriangular
quantum universal enveloping superalgebra $(U_{h}(\mathfrak{g}),R)$ which is
the first EK quantization.

Functoriality is the main feature in the proof of Theorem 35: There ia an
isomorphism of the first and the second EK quantizations of quasitriangular
superbialgebra as Hopf algebras.

And at last the Theorem 37: The quantization of a finite dimensional
superbialgebra commutes with taking the double $D(U_{h}(\mathfrak{g}))\cong
U_{h}(D(\mathfrak{g}))$

Next step (Section 9) is to show that (both of) EK quantizations are
isomorphic to the Drinfeld-Jimbo quantization. Following \cite{EK3} it
requires to show that the EK quantization is given by the desired generators
and relations. The main effort in this part is to prove the additional
quantum Serre-type relations which appear in the type $A-G$ superalgebras.
Since there are no such additional relations for $\mathfrak{gl}(1|1)$
superalgebra we omit this details.

The conclusive steps which lead to the main Theorem 48 were already
described above and do not require details.

We remark that the Geer's proof has no any restriction explicitly related to
semisimplicity of superalgebra and to semisimplicity of the category of its
finite dimensional modules therefore can be applied to non-semisimple
superalgebra $\mathfrak{gl}(1|1)$. The only feature one should be careful
about is that the tensor functors $F,\overline{F}$ between categories are
well defined for $\mathfrak{gl}(1|1)$ and their construction covers the set
of objects of our category of $\mathfrak{gl}(1|1)$ modules. We argued above
why it is the case.

This concludes the proof of our Theorem \ref{thm7} claiming braided tensor equivalence of the Drinfeld
category $\mathcal{D}$ and the category $\mathcal{C}_{\kappa }$ of
Drinfeld-Jimbo quantized superalgebra $\mathfrak{gl}(1|1)$.


Summarizing, we have checked that all the steps of the proof of braided
tensor equivalence in \cite{Gee} can be applied to the superalgebra $%
\mathfrak{gl}(1|1)$. It is based on the twist (\ref{FcDef}), which exists
and is unique, at least on the categories of the solutions of KZ equations
we consider. Unfortunately the formula (\ref{FcDef}) for twist is not
practically useful in explicit calculations because it requires to know the
explicit form of associator.

\section{Outlook}

\label{section6}

The proved braided tensor equivalence of non semisimple categories of $%
\mathcal{A}_{\Omega ,\kappa }$ and $U_{h}(\mathfrak{g})$ modules at generic
values of $\kappa $ is a preliminary step towards an understanding of
relation between corresponding modules for non generic values of $\kappa $.
In this case the problem actually becomes about a correspondence between the
categories of modules of logarithmic vertex operator superalgebra $V(%
\mathfrak{gl}(1|1),\kappa )$ and quantum group $U_{q}(\mathfrak{gl}(1|1))$.
Despite a\ big progress done in understanding of this correspondence in the
last years for non-superalgebra case, the situation with superalgebras
remains, to our knowledge, unclear. Recall that in the known cases of such
correspondence for non superalgebras the relevant second partner of the
correspondence is restricted quantum group, or in the case of logarithmic
VOA, unrolled restricted quantum group \cite{CMR}, \cite{Ru}. As we
mentioned in Subsection 2.1 an essential progress has been achieved recently
in \cite{CMY} in understanding of vertex tensor category structure of $V(%
\mathfrak{gl}(1|1),\kappa )$ for any $\kappa $ including non-generic $\kappa
$. It would be interesting to understand what is the quantum group partner
for $V(\mathfrak{gl}(1|1),\kappa )$ - modules category for non-generic
values of $\kappa $. On a VOA part of the correspondence a rigorous
construction of intertwining operators for vertex operator superalgebras at
non-generic $\kappa $ is an important first step (for non-superalgebras it
was recently done in \cite{McR}). Another hard problem is to understand
practical applicability of vertex tensor categories structures (see \cite%
{HuRev} and references therein) in concrete cases of superalgebras \cite%
{CKMcR}.

Another interesting problem is a logarithmic generalization of the way to
construct all the solutions of KZ equations for corrtelation function
including non-semisimple finitely generated modules, by an integration
operator as in (\ref{integr}) from some minimal set of basic solutions. It
is natural to expect as a result a sort of logarithmic deformations of
hypergeometric functions structures discovered in \cite{ScheVar}.

\section{Appendix A}

\label{appA}

In this Appendix we collect some data about $\mathfrak{gl}(1|1)$ and details of solutions
of its KZ equations.

\subsection{Asymptotic solutions of KZ equation}

\label{section:subAsymp} 

\begin{lemma}
\label{lemma: A1} If there is an eigenvector (not generalized) $v$ of $%
\Omega _{12}$ with eigenvalue $\lambda $, and there are no eigenvalues of $%
\Omega _{12}$ such that $\lambda +n\kappa ,n\in
\mathbb{N}
$, then there exists unique asymptotic solution around $x=0$
\begin{equation*}
f(x)=x^{\lambda /\kappa }(v+o(x)),\;\;\lim_{x\rightarrow 0}o(x)=0
\end{equation*}
\end{lemma}

\textit{Proof. }By not generalized eigenvector we mean that $v$ is not a
member of a Jordan block. We check existence and uniqueness of asymptotic
solution of the form
\begin{equation}
f(x)=x^{\lambda /\kappa
}(v+xv_{1}+x^{2}v_{2}+...),\;\;o(v)=\sum_{n=1}x^{n}v_{n}  \label{as}
\end{equation}%
with some perhaps infinite set of vectors $v_{n}$. After the substitution of
it into the left hand side of the equation (\ref{KZf}) we get%
\begin{equation}
lhs=x^{\lambda /\kappa }[\lambda x^{-1}v+(\lambda +\kappa )v_{1}+x(\lambda
+2\kappa )v_{2}+x^{2}(\lambda +3\kappa )v_{3}+...]  \label{lhs}
\end{equation}%
The right hand side we rewrite in the vicinity of $x=0$ as
\begin{equation*}
\frac{\Omega _{12}}{x}-\Omega _{23}(1+x+x^{2}+...)
\end{equation*}%
and now we act by it onto (\ref{as}):%
\begin{eqnarray}
rhs &=&x^{\lambda /\kappa }\left( \frac{\Omega _{12}}{x}-\Omega
_{23}(1+x+x^{2}+...)\right) (v+xv_{1}+x^{2}v_{2}+...)  \label{rhs} \\
&=&x^{\lambda /\kappa }[\lambda vx^{-1}+(\Omega _{12}v_{1}-\Omega
_{23}v)+(\Omega _{12}v_{2}-\Omega _{23}v_{1}-\Omega _{23}v)x  \notag \\
&&+(\Omega _{12}v_{3}-\Omega _{23}v_{2}-\Omega _{23}v_{1})x^{2}+...]  \notag
\end{eqnarray}%
Now we compare the multipliers of the same powers of $x$ in (\ref{lhs}) and (%
\ref{rhs}) and get the infinite set of equations
\begin{eqnarray}
x^{0} &:&(\Omega _{12}-(\lambda +\kappa )Id)v_{1}=\Omega _{23}v,
\label{rhs1} \\
x^{1} &:&(\Omega _{12}-(\lambda +2\kappa )Id)v_{2}=\Omega _{23}v_{1}+\Omega
_{23}v,  \notag \\
x^{2} &:&(\Omega _{12}-(\lambda +3\kappa )Id)v_{3}=\Omega _{23}v_{2}+\Omega
_{23}v_{1},  \notag \\
&&.......  \notag
\end{eqnarray}%
They can be solved one after another. Indeed, the right hand side of the
first equation is a known vector. $\det [\Omega _{12}-(\lambda +\kappa
)Id]\neq 0$ because $\lambda +\kappa $ is not an eigenvalue of $\Omega _{12}$%
. Therefore the first equation has a unique solution $v_{1}$. The same
arguments can now be applied to the second equation : $\Omega _{23}v_{1}$ is
now a known vector. We can solve the second equation for $v_{2}$, which is
possible because $\det [\Omega _{12}-(\lambda +2\kappa )Id]\neq 0$, for $%
\lambda +2\kappa $ is not an eigenvalue of $\Omega _{12}$. And so on. Thus
we find uniquely each vector $v_{i}$ by this recurrent procedure, which
proves the statement. We don't discuss the convergency question of the
infinite sum of vectors in $o(v)$ because we prove only the existence of
asymptotic expansion.\ \

$\square $

The case of a Jordan block requires more general ansatz. The operator $%
x^{\Omega _{12}/\kappa }$ is a well defined operator on any finite
dimensional representation space on which $\Omega _{12}$ acts nilpotently.
In this case the operator
\begin{equation}
x^{\Omega _{12}/\kappa }=\sum_{i=0}^{n}\frac{(\ln x)^{i}}{i!}\frac{\Omega
_{12}^{i}}{\kappa ^{i}}  \label{logex}
\end{equation}%
where $n$ is the degree of nilpotency of $\Omega _{12}$. Then we can
reformulate the lemma in the following way.

\begin{lemma}
\label{lemma: A2} If there is a Jordan block of $\Omega _{12}$ with
eigenvalue $\lambda $ with the set of eigenvectors $v^{(i)},i=0,...,n-1$, $%
\Omega _{12}v^{(i)}=\lambda v^{(i)}+v^{(i-1)}$, ($v^{(-1)}=0$) and there are
no eigenvalues of $\Omega _{12}$ such that $\lambda +n\kappa ,n\in
\mathbb{N}
$, then there exist $n$ asymptotic solutions around $x=0$ of the form
\begin{equation}
f_{i}(x)=x^{\lambda /\kappa }(v^{(i)}(\ln x)^{i}+\kappa ^{-1}v^{(i-1)}(\ln
x)^{i-1}+o^{(i)}(x)),\;\;\lim_{x\rightarrow
0^{+}}o^{(i)}(x)=0,\;\;i=0,...,n-1  \label{asvj}
\end{equation}
\end{lemma}

\textit{Proof. }To make the presentation more clear we put $\kappa =1$ and
prove the statement for the case of rank $n=2$ Jordan block. With a more
lengthy formulas the same proof can be repeated for $n>2$. The claim of the
lemma for $f_{0}(x)$ becomes identical to the claim of the Lemma \ref{lemma:
A1}, with the same proof and the same form of the vector $%
o^{(0)}(x)=xv_{1}+x^{2}v_{2}+...$. Now we prove the lemma for $f_{1}(x).$ We
show existence and uniqueness of $v_{j},u_{j},j=1,2,...$ such that%
\begin{eqnarray}
f_{1}(x) &=&x^{\Omega _{12}}(v^{(1)}\ln x+v^{(0)}+o^{(1)}(x)),  \label{as1}
\\
o^{(1)}(x) &=&\sum_{j=1}^{\infty }v_{j}x^{j}\ln x+\sum_{j=1}^{\infty
}u_{j}x^{j}  \notag
\end{eqnarray}%
First we prove existence of the vectors $v_{j}$. We substitute this ansatz
for $o^{(1)}(x)$ into the KZ equation (\ref{KZf}). We see that the terms
proportional to $\ln x/x$ and $1/x$ cancel. Using the same expansion in
powers of $x$ of the term $\Omega _{23}/(x-1)$ in before and extracting the
terms containing $\ln x$ we get the equations
\begin{eqnarray}
\ln x &:&(\Omega _{12}-(\lambda +1)Id)v_{1}=\Omega _{23}v^{(0)},  \label{as2}
\\
x\ln x &:&(\Omega _{12}-(\lambda +2)Id)v_{2}=\Omega _{23}(v^{(0)}+v_{1})
\notag \\
&&........  \notag
\end{eqnarray}%
As before we can solve these equations for $v_{1},v_{2},...$ sequentially
because $\lambda +n,n\geq 1$ is not an eigenvalue of $\Omega _{12}$ and the
right hand side of these equations are known vectors. After we found $v_{i}$%
s we do the same extracting on both hand side of KZ equation the terms which
are not proportional to $\ln x$. We get%
\begin{eqnarray}
x &:&(\Omega _{12}-(\lambda +1)Id)u_{1}=\Omega _{23}v^{(1)}+v_{1},
\label{as3} \\
x^{2} &:&(\Omega _{12}-(\lambda +2)Id)u_{2}=\Omega _{23}v^{(1)}+v_{2}+\Omega
_{23}u_{1},  \notag \\
&&........  \notag
\end{eqnarray}%
By the same reasons as before the equations can be uniquely solved
sequentially for $u_{i}$. This completes the proof.

$\square $

In the same way we can prove similar statements about existence of unique
asymptotic solutions of the \ref{KZf} equation around $x=1,x<1$.

\begin{lemma}
\label{lemma: A3} If there is an eigenvector $v$ of $\Omega _{23}$ with
eigenvalue $\lambda $, and there are no eigenvalues of $\Omega _{23}$ such
that $\lambda +n\kappa ,n\in
\mathbb{N}
$, then there exists unique asymptotic solution around $x=1$ of the form
\begin{equation}
f(x)=(1-x)^{-\lambda /\kappa }(v+o(x)),\;\;\lim_{x\rightarrow 1^{-}}o(x)=0
\label{x1}
\end{equation}%
in the case this eigenvector is not a member of a Jordan block. For the case
of Jordan block of the size $n$ the $n$ asymptotic solutions are of the form
\begin{eqnarray*}
f_{i}(x) &=&(1-x)^{-\lambda /\kappa }(v^{(i)}(\ln (1-x))^{i}+\kappa
^{-1}v^{(i-1)}(\ln (1-x))^{i-1}+o^{(i)}(x)),\;\; \\
\lim_{x\rightarrow 1^{-}}o^{(i)}(x) &=&0,\;\;i=0,...,n-1
\end{eqnarray*}
\end{lemma}

Proof is the same as for Lemmas \ref{lemma: A1},\ref{lemma: A2}.

\begin{corol}
\label{corol: A4}
If the above restriction conditions on the parameters of
typical modules are satisfied an equivalent form of asymptotic solutions of (%
\ref{KZf}) around $x=0$ is
\begin{equation}
f(x)=x^{\Omega _{12}/\kappa }(v_{b}+o(v))  \label{op}
\end{equation}%
where $v_{t}$ is the same as $v$ in the case when there are no Jordan block
structure for the action of $\Omega _{12}$, and $v_{b}$ is the bottom vector
$v^{(n-1)}$ when there is a Jordan block of size $n$ for the action of $%
\Omega _{12}$.
\end{corol}

\textit{Proof.} In the case without Jordan block this is just change of
notations. In the case when there is Jordan block of size $n$ we split $%
\Omega _{12}=$ $\Omega _{12}^{d}+\Omega _{12}^{nil}$ into diagonal and
nilpotent parts and write $x^{\Omega _{12}/\kappa }=x^{\Omega
_{12}^{d}/\kappa }\sum_{i}\frac{1}{i!}\left( \frac{\Omega _{12}^{nil}}{%
\kappa }\ln x\right) ^{i}$. The action of it on the bottom vector of the set
of generalized eigenvectors of $\Omega _{12}$ will generate the sum of
vectors proportional to $(\ln x)^{i}v^{(i)}$ where $v^{(i)}$ are the same as
in (\ref{asvj}). Therefore the representation (\ref{asvj}) is related to the
expansion (\ref{op}) by a change of basis of solutions of KZ equation.

$\square $

This corollary enables to use without changes the standard proofs of BTC
structure of category of $\mathfrak{gl}(1|1)$-modules with associator and
braiding defined through the KZ solutions and their monodromies.

\subsection{Basis for $\mathfrak{gl}(1|1)$ and its modules}

\label{subgl11}

The $\mathfrak{gl}(1|1)$ generators are $E,N,$ $\psi ^{\pm }$ with
commutation relations $[N,\psi ^{\pm }]=\pm \psi ^{\pm }$, $\{\psi ^{+},\psi
^{-}\}=E$ and $E$ is central. (Maybe some other choice of basis will be more
convenient?) Chevalley involution can be chosen as $\omega (E)=-E,$ $\omega
(N)=-N,$ $\omega (\psi ^{\pm })=\pm \psi ^{\mp }$ and produces the dual
representation. The basis for typical representation $\mathcal{T}_{e,n}$ of $%
gl(1|1)$ can be chosen as
\begin{equation}
N=\left(
\begin{array}{cc}
n+1/2 & 0 \\
0 & n-1/2%
\end{array}%
\right) ,\;E=\left(
\begin{array}{cc}
e & 0 \\
0 & e%
\end{array}%
\right) ,\;\psi ^{+}=\left(
\begin{array}{cc}
0 & e \\
0 & 0%
\end{array}%
\right) ,\;\psi ^{-}=\left(
\begin{array}{cc}
0 & 0 \\
1 & 0%
\end{array}%
\right)  \label{glbt}
\end{equation}%
The basis for weights of module $\mathcal{T}_{e,n}$ is $u=\uparrow =\binom{1%
}{0}$ (even highest weight), and $v=\downarrow =\binom{0}{1}$ (odd), and for
dual module $\mathcal{T}_{e,n}^{\ast }$ -- $u^{\ast }=\binom{0}{1}$ (odd
lowest weight), and $v^{\ast }=\binom{-1}{0}$ (even). For one dimensional
atypical representation $\mathcal{A}_{n}$ there is one vector $v_{0}$ with
the action of the algebra generators $\psi ^{+}v_{0}=\psi ^{-}v_{0}=Ev_{0}=0$%
, $Nv_{0}=nv_{0}$. The algebra action on it explicitly:%
\begin{equation}
N\cdot \uparrow =(n+1/2)\uparrow ,\;N\cdot \downarrow =(n-1/2)\uparrow
,\;\psi ^{+}\cdot \uparrow =\psi ^{-}\cdot \downarrow =0,\;\psi ^{-}\cdot
\uparrow =\downarrow ,\;\psi ^{+}\cdot \downarrow =e\uparrow  \label{acttyp}
\end{equation}%
For four dimensional atypical representation $\mathcal{P}_{n}$ one can
choose
\begin{eqnarray}
N &=&\left(
\begin{array}{cccc}
n+1 & 0 & 0 & 0 \\
0 & n & 0 & 0 \\
0 & 0 & n & 0 \\
0 & 0 & 0 & n-1%
\end{array}%
\right) ,\;\psi ^{+}=\frac{1}{2}\left(
\begin{array}{cccc}
0 & 1 & 1 & 0 \\
0 & 0 & 0 & 1 \\
0 & 0 & 0 & -1 \\
0 & 0 & 0 & 0%
\end{array}%
\right) ,\;\psi ^{-}=\frac{1}{2}\left(
\begin{array}{cccc}
0 & 0 & 0 & 0 \\
-1 & 0 & 0 & 0 \\
1 & 0 & 0 & 0 \\
0 & 1 & 1 & 0%
\end{array}%
\right) ,  \label{glbp} \\
E &=&0\times Id_{4}  \notag
\end{eqnarray}%
And the weights of the module
\begin{equation}
u_{1}=t=\left(
\begin{array}{c}
0 \\
1 \\
1 \\
0%
\end{array}%
\right) ,\;v_{1}=r=\left(
\begin{array}{c}
1 \\
0 \\
0 \\
0%
\end{array}%
\right) ,\;v_{2}=l=\left(
\begin{array}{c}
0 \\
0 \\
0 \\
1%
\end{array}%
\right) ,\;u_{2}=b=\frac{1}{2}\left(
\begin{array}{c}
0 \\
1 \\
-1 \\
0%
\end{array}%
\right)  \label{glpv}
\end{equation}%
the even vectors are $u_{1,2}$, the odd \ $v_{1,2}$. This module is self
dual. The algebra action on it%
\begin{align}
N\cdot t& =nt,\;N\cdot r=(n+1)r,\;N\cdot l=(n-1)l,\;N\cdot b=nb,
\label{actproj} \\
\psi ^{+}\cdot t& =r,\;\psi ^{+}\cdot l=b,\;\psi ^{+}\cdot r=\psi ^{+}\cdot
b=0,\;  \notag \\
\psi ^{-}\cdot t& =l,\;\psi ^{-}\cdot r=-b,\;\psi ^{-}\cdot l=\psi ^{-}\cdot
b=0,\;  \notag
\end{align}

We will use the following choice of Casimir element
\begin{equation}
\Omega=NE+EN+\psi^{-}\psi^{+}-\psi^{+}\psi^{-}+E^{2}  \label{cas}
\end{equation}
and its tensor analog
\begin{equation}
\Omega_{ij}=N_{i}\otimes E_{j}+E_{i}\otimes N_{j}+\psi_{i}^{-}\otimes\psi
_{j}^{+}-\psi_{i}^{+}\otimes\psi_{j}^{-}+E_{i}\otimes E_{j}  \label{cast}
\end{equation}
where the lower indices denote the spaces where the generator acts.

$\widehat{\mathfrak{gl}}(1|1)$ commutation relations
\begin{equation}
\lbrack N_{r},E_{s}]=rk\delta _{r+s},\;[N_{r},\psi _{s}^{\pm }]=\pm \psi
_{r+s}^{\pm },\;\{\psi _{r}^{+},\psi _{s}^{-}\}=E_{r+s}+rk\delta _{r+s}
\label{acm}
\end{equation}%
One can rescale generators in such a way that $k$ will become 1 (if it is
not 0), but we will keep it. The generic $k$ will mean $e/k\notin
\mathbb{Z}
$ for all the modules involved into correlation function, as well as for all
the modules appearing in tensor product decomposition. A remark: the
structure of all modules for non generic $k$ for $\widehat{\mathfrak{gl}}%
(1|1)$ and their tensor product decomposition is of course well known, but
the KZ for this case and its solutions is another (next...) problem.

Conformal dimension of Virasoro primary field $h=e\left( n+\frac{e}{2}%
\right) $.

We are going to find basis for invariants of level zero KZ equations for $%
N=2,3,4$. Recall that level zero equations in the case of $\mathfrak{gl}%
(1|1) $ means that $\sum e_{i}=0$, if typical representations are involved in correlation
function. In addition the invariants can be classified according to the $N$%
-grading of the space of states $V$ of correlation function.

\subsection{Examples of solutions of KZ equation for correlation functions}

\label{subCorf}

In this section we collect examples of explicit form of KZ $N=2,3$ solutions
on the space of $\mathfrak{gl}(1|1)$ invariant functions. This class of
solutions is the most interesting in the context of KZ equations for
correlation functions of intertwining operators of affine Lie superalgebra $%
\mathfrak{gl}(1|1)^{\vee }$. Similar calculations has been done in the paper
\cite{Troo}.

\textbf{1. }$N=2$

There is one invariant for $\mathcal{TT}$ correlation function in the basis
described above $I_{0}^{\mathcal{TT}}=\uparrow \downarrow +\downarrow
\uparrow ,$and the list of invariants for $\mathcal{PP}$ correlation
function is
\begin{align}
I_{-1}^{\mathcal{PP}}& =rb-br  \label{PPinv} \\
I_{0,1}^{\mathcal{PP}}& =tb+rl-lr+bt  \notag \\
I_{0,2}^{\mathcal{PP}}& =bb  \notag \\
I_{1}^{\mathcal{PP}}& =lb-bl  \notag
\end{align}%
The first subindex denotes the value of $n_{1}+n_{2}$. (Recall that it is
not an eigenvalue of $N$ acting on the tensor product state. The latter is 0
for $\mathfrak{g}$-invariant correlation function.) Projection of KZ $N=2$
equation onto this basis gives an ODE with solutions%
\begin{equation}
f(z_{1},z_{2})=[A(z_{1}-z_{2})^{\delta _{12}/k}]I_{0}^{\mathcal{TT}%
},\;\delta _{ij}=n_{i}e_{j}+n_{j}e_{i}+e_{i}e_{j}  \label{sol1}
\end{equation}%
for $\mathcal{T}_{e_{1},n_{1}}\mathcal{T}_{-e_{1},n_{2}}$ correlation
function ($A$ is a constant), and solutions
\begin{eqnarray}
f(z_{1},z_{2}) &=&const\times I_{\pm 1}^{\mathcal{PP}},\text{ for }%
n_{1}+n_{2}=\pm 1  \label{sol2} \\
f(z_{1},z_{2}) &=&AI_{0,2}^{\mathcal{PP}}+(2A\kappa ^{-1}\ln
(z_{1}-z_{2})+B)I_{0,1}^{\mathcal{PP}}\text{ \ for }n_{1}+n_{2}=0  \notag
\end{eqnarray}%
where $A,B$ are constants. This is an example of logarithms in correlation
functions of logarithmic vertex operator algebras.

\textbf{2. }$N=3$

There are two invariants for $\mathcal{TTT}$ correlation in the same
notations as above

\begin{align}
I_{-1/2}^{\mathcal{TTT}}& =(\uparrow \uparrow \downarrow +\uparrow
\downarrow \uparrow +\downarrow \uparrow \uparrow )  \label{TTTinv} \\
I_{+1/2}^{\mathcal{TTT}}& =(e_{1}\uparrow \downarrow \downarrow
-e_{2}\downarrow \uparrow \downarrow +e_{3}\downarrow \downarrow \uparrow ),
\notag
\end{align}%
(Of course $e_{1}+e_{2}+e_{3}=0$.) Invariants of $\mathcal{TTP}$
correlations are%
\begin{align}
I_{-1}^{\mathcal{TTP}}& =\uparrow \uparrow b-\uparrow \downarrow
r-\downarrow \uparrow r  \label{TTPinv} \\
I_{0,1}^{\mathcal{TTP}}& =e_{1}(\uparrow \uparrow l+\uparrow \downarrow
t+\downarrow \uparrow t)+\uparrow \downarrow b+\downarrow \downarrow r
\notag \\
I_{0,2}^{\mathcal{TTP}}& =\uparrow \downarrow b+\downarrow \uparrow b  \notag
\\
I_{1}^{\mathcal{TTP}}& =e_{1}(\uparrow \downarrow l+\downarrow \uparrow
l)+\downarrow \downarrow b  \notag
\end{align}%
and the list of invariants of $\mathcal{PPP}$ correlations are
\begin{align}
I_{-2}^{\mathcal{PPP}}& =rrb-rbr+brr  \label{PPPinv} \\
I_{-1,1}^{\mathcal{PPP}}& =trb-tbr-rrl-rbt+lrr+brt  \notag \\
I_{-1,2}^{\mathcal{PPP}}& =rtb+rrl-rlr+rbt-btr-brt  \notag \\
I_{-1,3}^{\mathcal{PPP}}& =rbb-brb  \notag \\
I_{-1,4}^{\mathcal{PPP}}& =rbb-bbr  \notag \\
I_{0,1}^{\mathcal{PPP}}& =btb+brl-blr+bbt  \notag \\
I_{0,2}^{\mathcal{PPP}}& =bbb  \notag \\
I_{1,1}^{\mathcal{PPP}}& =tlb-tbl-rll+llr-lbt+blt  \notag \\
I_{1,2}^{\mathcal{PPP}}& =ltb+lrl-llr+lbt-btl-blt  \notag \\
I_{1,3}^{\mathcal{PPP}}& =lbb-bbl  \notag \\
I_{1,4}^{\mathcal{PPP}}& =blb-bbl  \notag \\
I_{2}^{\mathcal{PPP}}& =llb-lbl+bll  \notag
\end{align}%
Projection of KZ equation in the form (\ref{KZf}) onto these bases gives
systems of ODEs with the following solutions. If the space of invariants
with fixed first subindex, i.e. fixed sum of $n_{1}+n_{2}+n_{3}$ is one
dimensional equal to $I$ then the solution for correlation function in all
three cases can be written as
\begin{equation}
f(x)=Ax^{\alpha /\kappa }(1-x)^{\beta /\kappa }I  \label{sol31}
\end{equation}%
where $A\in
\mathbb{C}
$ is a constant, and $\alpha ,\beta $ are eigenvalues of $\Omega
_{12},\Omega _{23}$ acting on $I$ respectively.

In the $\mathcal{TTP}$ case with $n_{1}+n_{2}+n_{3}=0$ solution contains
logarithms:%
\begin{equation}
f(x)=Ax^{\delta _{12}/\kappa }(1-x)^{\delta _{23}/\kappa }[I_{0,1}^{\mathcal{%
TTP}}+(B+\frac{e_{1}}{\kappa }(\ln (1-x)-\ln x))I_{0,2}^{\mathcal{TTP}}]
\label{TTP1}
\end{equation}

In the $\mathcal{PPP}$ case with $n_{1}+n_{2}+n_{3}=0$ the solution is
trivial
\begin{equation}
f(x)=AI_{0,1}^{\mathcal{PPP}}+BI_{0,2}^{\mathcal{PPP}},\;A,B\in
\mathbb{C}
\label{PPP0}
\end{equation}%
But in the case $n_{1}+n_{2}+n_{3}=\pm 1$ there are logarithms in the
solutions:%
\begin{eqnarray}
f^{\pm }(x) &=&A^{\pm }I_{\pm 1,1}^{\mathcal{PPP}}+B^{\pm }I_{\pm 1,2}^{%
\mathcal{PPP}}+\left( C_{3}^{\pm }+\frac{A^{\pm }-B^{\pm }}{\kappa }\ln x+%
\frac{B^{\pm }-2A^{\pm }}{\kappa }\ln (1-x)\right) I_{\pm 1,3}^{\mathcal{PPP}%
}  \notag \\
&&+\left( C_{4}^{\pm }+\frac{B^{\pm }}{\kappa }\ln x+\frac{A^{\pm }-B^{\pm }%
}{\kappa }\ln (1-x)\right) I_{\pm 1,3}^{\mathcal{PPP}}  \label{PPP1}
\end{eqnarray}%
where $A^{\pm },B^{\pm },C_{3,4}^{\pm }$ are constants.

Another interesting problem is structure of solutions of KZ equations on a
wider $N$-graded spaces, not necessarily invariants of $\mathfrak{gl}(1|1)$.
We will address this problem elsewhere.

\section{Appendix B}

\label{appB}

Here we will describe the basis and tensor product decomposition of $U_{h}(%
\mathfrak{gl}(1|1))$-modules and will prove the Proposition 3.

We will choose $i\pi \kappa ^{-1}=h$ and consider real $\kappa $. We use the
following matrix basis for the three types of $U_{h}(\mathfrak{gl}(1|1))$%
-modules $\mathcal{T}_{e,n}^{\kappa },\mathcal{A}_{n}^{\kappa },\mathcal{P}%
_{n}^{\kappa }$ included into $\mathcal{C}_{\kappa }$, as the basis for
construction of tensor ring. For $\mathcal{T}_{e,n}^{\kappa }$%
\begin{equation*}
E=\left(
\begin{array}{cc}
e & 0 \\
0 & e%
\end{array}%
\right) ,\;N=\left(
\begin{array}{cc}
n+1/2 & 0 \\
0 & n-1/2%
\end{array}%
\right) ,\;\psi ^{+}=\left(
\begin{array}{cc}
0 & 2\sinh (eh) \\
0 & 0%
\end{array}%
\right) ,\;\;\psi ^{-}=\left(
\begin{array}{cc}
0 & 0 \\
1 & 0%
\end{array}%
\right)
\end{equation*}%
with the vectors of the module
\begin{equation*}
|e,n\rangle =\left(
\begin{array}{c}
1 \\
0%
\end{array}%
\right) \text{ \ (even)},\;|e,n-1\rangle =\psi ^{-}|e,n\rangle =\left(
\begin{array}{c}
0 \\
1%
\end{array}%
\right) \text{ \ (odd)}
\end{equation*}
and for four dimensional module we choose
\begin{eqnarray*}
N &=&\left(
\begin{array}{cccc}
n+1 & 0 & 0 & 0 \\
0 & n & 0 & 0 \\
0 & 0 & n & 0 \\
0 & 0 & 0 & n-1%
\end{array}%
\right) ,\;\psi ^{+}=\left(
\begin{array}{cccc}
0 & 1 & -e^{h} & 0 \\
0 & 0 & 0 & e^{h} \\
0 & 0 & 0 & 1 \\
0 & 0 & 0 & 0%
\end{array}%
\right) , \\
\psi ^{-} &=&\left( \allowbreak
\begin{array}{cccc}
0 & 0 & 0 & 0 \\
-1 & 0 & 0 & 0 \\
-e^{-h} & 0 & 0 & 0 \\
0 & e^{-h} & -1 & 0%
\end{array}%
\right) ,\;E=0\times Id_{4},
\end{eqnarray*}%
The coordinates of the vectors of the four dimensional vector space of this
representation are graded as in (\ref{glpv}). Let us note that there are
many other matrix presentations of $\mathcal{P}_{n}^{\kappa }$ module which
can contain some more free numerical parameters.

\textit{Proof of Proposition 3}. With these basis we can consider
decomposition of tensor product of this set of three types of modules using
the coproduct (\ref{qcop}) and show that under some suitable assumptions on
parameters of modules they form a ring. The cases
\begin{equation*}
\mathcal{A}_{n}^{\kappa }\otimes \mathcal{A}_{n^{\prime }}^{\kappa }=%
\mathcal{A}_{n+n^{\prime }}^{\kappa },\;\mathcal{A}_{n}^{\kappa }\otimes
\mathcal{T}_{e,n^{\prime }}^{\kappa }=\mathcal{T}_{e,n+n^{\prime }}^{\kappa
},\;\mathcal{A}_{n}^{\kappa }\otimes \mathcal{P}_{e,n^{\prime }}^{\kappa }=%
\mathcal{P}_{e,n+n^{\prime }}^{\kappa }
\end{equation*}%
are obvious. More interesting are the remaining three cases.

Consider $\mathcal{T}_{e,n}^{\kappa }\otimes \mathcal{T}_{e^{\prime
},n^{\prime }}^{\kappa }$. The calculations of tensor product decomposition
of two $U_{ih}(\mathfrak{gl}(1|1))$-modules $\mathcal{T}_{e_{1},n_{1}}^{%
\kappa }\otimes \mathcal{T}_{e_{2},n_{2}}^{\kappa }$ is completely parallel
to the same calculations for $\mathfrak{gl}(1|1)$-modules. $\mathcal{T}%
_{e,n}^{\kappa }$ has two states - the highest weight $v_{1}=|\uparrow
\rangle $ Grassmann even and $v_{2}=\psi ^{-}v_{1}=|\downarrow \rangle $ -
Grassmann odd. We can start from two vectors $w_{2}=\alpha _{2}|\uparrow
\rangle \otimes |\downarrow \rangle +\beta _{2}|\downarrow \rangle \otimes
|\uparrow \rangle $ and $u_{1}=\alpha _{1}|\uparrow \rangle \otimes
|\downarrow \rangle +\beta _{1}|\downarrow \rangle \otimes |\uparrow \rangle
$ with constraint $\alpha _{2}=-\beta _{1}\beta _{2}/\alpha _{1}$ which
guarantees their orthogonality. We consider $w_{2}$ as highest weight of a
grading reversed module, i.e. $\Delta (\psi ^{+})w_{2}=0$. It gives $\beta
_{2}=-2\alpha _{2}e^{-he_{1}}\sinh (e_{2}h)$. And we consider $u_{1}$ \ as
lowest weight module, with Grassmann even highest weight. It means $\Delta
(\psi ^{-})u_{1}=0$, which gives $\beta _{1}=\alpha _{1}e^{he_{2}}$. Then
one can easily check that corresponding lowest weight module of the first
(grading reversed) module is $2\alpha _{1}\sinh ((e_{1}+e_{2})h)|\downarrow
\rangle \otimes |\downarrow \rangle $, and highest weight of the second
module is $\alpha _{2}\sinh ((e_{1}+e_{2})h)/\sinh (e_{1}h)|\uparrow \rangle
\otimes |\uparrow \rangle $. We see that conditions $\sinh
((e_{1}+e_{2})h)\neq 0$, $\sinh (e_{1}h)\neq 0$, which mean $e_{1}/\kappa
\notin
\mathbb{Z}
$, $(e_{1}+e_{2})/\kappa \notin
\mathbb{Z}
\backslash \{0\}$ are sufficient for decomposition
\begin{equation*}
\mathcal{T}_{e_{1},n_{1}}^{\kappa }\otimes \mathcal{T}_{e_{2},n_{2}}^{\kappa
}=\mathcal{T}_{e_{1}+e_{2},,n_{1}+n_{2}+1/2}^{\kappa }\oplus \mathcal{T}%
_{e_{1}+e_{2},,n_{1}+n_{2}-1/2}^{\kappa \prime }
\end{equation*}%
In the case $e_{1}+e_{2}=0$ one can check that any vector of the form $%
|t\rangle =\alpha |\uparrow \rangle \otimes $ $|\downarrow \rangle +\beta
|\downarrow \rangle \otimes $ $|\uparrow \rangle $ with $\alpha \neq
e^{he_{1}}\beta $ serves as the $|t\rangle $-vector in the basis of the $%
\mathcal{P}_{n_{1}+n_{2}}^{\kappa }$ module of four vectors of the tensor
product $\mathcal{T}_{e_{1},n_{1}}^{\kappa }\otimes \mathcal{T}%
_{-e_{1},n_{2}}^{\kappa }$. We see that the tensor product ring composed of
the $U_{h}(\mathfrak{gl}(1|1))$-modules $\mathcal{A}_{n}^{\kappa },\mathcal{T%
}_{e,n}^{\kappa },\mathcal{P}_{n}^{\kappa }$ is the same as the tensor
product ring of the category $\mathcal{C}_{\kappa }$ composed of $\mathcal{A}%
_{n},\mathcal{T}_{e,n},\mathcal{P}_{n}$ for restriction on parameters the
same as in the Proposition 3.

$\square $

\end{document}